\documentclass[twocolumn,prd,amsmath,amssymb,aps,floatfix,10pt,superscriptaddress,nofootinbib]{revtex4-2}

\usepackage{graphicx}
\usepackage{dcolumn}
\usepackage{bm}
\usepackage{hyperref}
\usepackage{xcolor}
\usepackage{natbib}
\usepackage[compact]{titlesec}
\usepackage[title]{appendix}
\usepackage[normalem]{ulem}

\titlespacing{\section}{2pt}{*2}{*1}
\titlespacing{\subsection}{2pt}{*2}{*1}
\titlespacing{\subsubsection}{2pt}{*2}{*1}

\makeatletter
\newcommand\footnoteref[1]{\protected@xdef\@thefnmark{\ref{#1}}\@footnotemark}
\makeatother

\newcommand{\victor}[1]{}

\begin{document}
    
    \title{The broadening of universal relations at the birth and death of a neutron star}
    
    \author{Victor Guedes}
    \email{tpx5df@virginia.edu}
    \affiliation{Department of Physics, University of Virginia, Charlottesville, VA 22904, USA}

    \author{Shu Yan Lau}
    \affiliation{Department of Physics, University of Virginia, Charlottesville, VA 22904, USA}

    \author{Cecilia Chirenti}
    \affiliation{Department of Astronomy, University of Maryland, College Park, MD 20742, USA}
    \affiliation{Astroparticle Physics Laboratory, NASA/GSFC, Greenbelt, MD 20771, USA}
    \affiliation{Center for Research and Exploration in Space Science and Technology, NASA/GSFC, Greenbelt, MD 20771, USA}
    \affiliation{Center for Mathematics, Computation, and Cognition, UFABC, Santo André, SP 09210-170, Brazil}

    \author{Kent Yagi}
    \affiliation{Department of Physics, University of Virginia, Charlottesville, VA 22904, USA}
    
    \begin{abstract}

        Certain relations among neutron-star observables that are insensitive to the underlying nuclear matter equation of state are known to exist. Such universal relations have been shown to be valid for cold and stationary neutron stars. Here, we study these relations in more dynamic scenarios: protoneutron stars and hypermassive neutron stars, allowing us to investigate the time evolution of these relations from the birth to the death of a neutron star. First, we study protoneutron stars. We use an effective equation of state, extracted from three-dimensional core-collapse supernova simulations, to obtain the structure of spherically symmetric protoneutron stars. We then consider nonradial oscillations to compute their $f$-mode frequency ($f$), as well as  slow rotation and small tidal deformation, to compute their moment of inertia ($I$), spin-induced quadrupole moment ($Q$), and Love number. We find that well-established universal relations for cold neutron stars involving these observables (namely, the $I$-Love-$Q$ and $f$-Love relations) are approximately valid for protoneutron stars, with a deviation below $\approx$ 10$\%$ for a postbounce time above $\approx$ 0.5 s, considering eight different supernova progenitors and the SFHo equation of state. Next, we study hypermassive neutron stars. The bulk of a neutron star is defined as the region enclosed by the isodensity surface that corresponds to the maximum compactness ($C$) inside the star. We obtain a new universal relation between the $f$-mode frequency and the compactness of cold and nonrotating neutron stars, using bulk quantities. We show that this relation has an equation-of-state-variation of $\approx$ $3\%$, considering a set of ten equations of state. Bulk quantities of postmerger remnants can be obtained from numerical-relativity simulations. Using results from binary neutron star merger simulations, we study the evolution of hypermassive neutron stars on the $f$-$C$ plane, considering two different mass ratios and the SFHo equation of state. We find that the relation between the peak frequency of the gravitational-wave signal and the compactness from these hypermassive neutron stars deviates from the universal $f$-$C$ relation by 70 $-$ 80\%, when the peak frequency is taken directly as a proxy for the $f$-mode. Finally, we discuss the reliability of the universal relations in the context of future observations of gravitational waves from remnants of core-collapse supernovae or binary neutron star mergers.

    \end{abstract}
    
    \maketitle
    
    \section{Introduction}

        The structure of neutron stars (NSs) depends on their equation of state (EOS, the pressure as a function of the density). However, the extremely dense matter ($\gtrsim 2\times10^{14}$ g/cm$^{3}$) in the interior of these objects is not yet sufficiently constrained by experiments in nuclear/particle physics~\cite{foka2016}, thus making us turn to astrophysical observations to constrain their properties. Currently, we already have significant constraints on the EOS imposed by NS mass measurements through radio observations (PSR J0740+6620~\cite{cromartie}, PSR J0348+0432~\cite{antoniadis}, and PSR J1614-2230~\cite{fonseca}) of high-mass pulsars and NS radius measurements through X-ray observations (PSR J0030+0451~\cite{Miller:2019cac, Riley:2019yda, Raaijmakers:2019qny} and PSR J0740+6620~\cite{miller2021, riley2021, raaijmakers2021}). The recent detections of gravitational waves (GWs) from binary neutron star (BNS) mergers by the LIGO-Virgo collaboration, GW170817~\cite{ligo} and GW190425~\cite{ligo2}, have provided additional constraints~\cite{LIGOScientific:2018cki, LIGOScientific:2018hze, Chatziioannou:2020pqz} through measurements of the binary tidal deformability. Constraints on the EOS from multimessenger observations of NSs are found in, {\it e.g.}~\cite{Raaijmakers:2019dks, Dietrich:2020efo}. The moment of inertia of the primary pulsar in the double pulsar binary system PSR J0737-3039 has recently been constrained~\cite{Kramer:2021jcw}, and we also expect constraints on the EOS~\cite{lattimer} as the observation period increases and the measurement accuracy improves~\cite{LIGOScientific:2018hze}. Recently, the presence of high-frequency ($\sim$ kHz) quasiperiodic oscillations in short gamma-ray bursts was reported~\cite{chirenti2023}. These frequencies could be related to the oscillations of short-lived postmerger remnants (hypermassive NSs), and their future detection by third-generation ground-based GW detectors could provide strong constraints to the EOS~\cite{takami2014}.

        The EOS of NS matter can be constrained with forward modeling (see, {\it e.g.} the Bayesian analysis performed in~\cite{miller2020}). However, when modeling NSs, we have to assume that we know the EOS {\it a priori}, then we are able to compute NS observables, and finally use data from astrophysical observations and nuclear/particle physics experiments to obtain information about the EOS {\it a posteriori}. Many models for the EOS have been proposed, and this ``EOS freedom'' is one of the main sources of uncertainties in the theoretical modeling of these objects. 

        In general, the study of relations involving NS observables (mass, radius, tidal deformability, {\it etc.}) provides us with new tools to constrain the EOS. If the relation between two observables is strongly dependent on the EOS, we can constrain the EOS through measurements of both of the observables in the relation. On the other hand, if the relation is insensitive to the EOS, {\it i.e.} universal, having a measurement of one observable in the relation allows us to infer the other one even if such an observable is difficult to measure directly. One of the most well-studied universal relations are the ``$I$-Love-$Q$ relations'', proposed by Yagi and Yunes~\cite{yagi_2013_1, yagi_2013_2, yagi_2017}, which relate the moment of inertia ($I$), the spin-induced quadrupole moment ($Q$), and the tidal deformability ($\Lambda$, defined in terms of the Love number) of NSs.

        Several studies have extended the validity of $I$-Love-$Q$, {\it e.g.}: Maselli {\it et al.}~\cite{maseli} showed that the $I$-Love relation is universal throughout the inspiralling phase of a BNS system, depending on the inspiral frequency; Pappas and Apostolatos~\cite{pappas} argued that, more generally, the first four NS multipole moments are related in a way independent of the EOS, and such no-hair relations for NSs were studied in more detail both analytically~\cite{Stein:2013ofa} and numerically~\cite{Yagi:2014bxa}; additionally, the $I$-Love-$Q$ relations have also been studied for polytropic stars~\cite{benitez2021}, anisotropic stars~\cite{yagi_aniso}, hybrid stars~\cite{pasch}, strange stars~\cite{band}, incompressible stars~\cite{chan1, chan2}, boson stars~\cite{Adam:2022nlq}, dark stars~\cite{maselli_dark}, and gravastars~\cite{pani_grav, pani_grav_1, Uchikata:2016qku}; furthermore, they were also investigated in alternative theories of gravity~\cite{Doneva:2017jop,sham, pani_scalar, yagi_2013_1, yagi_2013_2, gupta_ilq, vylet2023, ajith}; possible explanations for this universality were reported by Yagi {\it et al.}~\cite{yagi_why}, and Sham {\it et al.}~\cite{sham_why}. Some studies, however, found cases in which the relations are not valid: Haskell {\it et al.}~\cite{haskell} showed that the universality is lost for long spin periods ($\gtrsim 10$ s) and strong magnetic fields ($\gtrsim 10^{12}$ G); Doneva {\it et al.}~\cite{doneva} found that the universality is lost for sequences of rapidly rotating NSs (up to the mass-shedding limit) when the rotational frequency is fixed, although the universality can be recovered when the dimensionless spin parameters are fixed, as shown by Pappas and Apostolatos~\cite{pappas} and Chakrabarti {\it et al.}~\cite{chakrabarti}.

        Newly-born protoneutron stars (PNSs), have also been shown to break the $I$-Love-$Q$ relations. PNSs are early remnants of core-collapse supernovae (CCSNe) and, although the supernova problem is not yet fully solved~\cite{burrows2021}, models of the stellar core after the bounce~\cite{burrows_lattimer, prakash1997, pons} allow us to investigate the validity of well-established universal relations for NSs during the early postbounce phase. Martinon {\it et al.}~\cite{martinon} showed that the $I$-Love-$Q$ relations for PNSs are different from the NS relations before $\sim 1$~s after the bounce, when the entropy gradients are significant. Marques {\it et al.}~\cite{marques2017} pointed out that, generally, the universality of the $I$-$Q$ relation is lost when thermal effects become important, even if the entropy is kept constant. Raduta {\it et al.}~\cite{raduta} confirmed the previous findings and demonstrated that the universality holds if thermodynamic conditions are kept the same, {\it i.e.} same entropy per baryon and lepton fraction. 
        
        Nonetheless, PNSs have been shown to follow some universal relations involving their quasinormal modes (QNMs). Torres-Forné {\it et al.}~\cite{torres} proposed relations between $f$-, $p$-, and $g$-modes and the average density of the shock region, as well as the surface gravity of the PNS; these relations were shown to not depend on the EOS, the neutrino treatment, or the progenitor mass. Sotani and collaborators~\cite{sotani2016, sotani2017, sotani2019_2, sotani2021_2, sotani2021_3} have proposed relations between the $f$-mode frequency and the average density of the PNS. This relation, in particular, has been shown to not depend on the EOS or the progenitor mass, and was also explored in recent works~\cite{mori2023, rodriguez2023}. Generally, these relations are not as tight as $I$-Love-$Q$ ({\it i.e.}, they show deviations larger than $\sim 1\%$), as a consequence of the extra degrees of freedom in the modeling of PNSs and the inherent errors of numerical simulations.

        QNMs of NSs have also been shown to correlate with other NS observables in an EOS-independent way. Andersson {\it et al.}~\cite{andersson1998} found an empirical relation between the $f$-mode frequency and the average density of the star, and proposed a similar relation for the damping time. Benhar {\it et al.}~\cite{benhar1999, benhar2004} extended these results to new EOSs and presented updated fits for these relations. Tsui and Leung~\cite{tsui2005} reported a relation between the $f$-mode frequency (multiplied by the NS mass) and the compactness of the star, the ``$f$-$C$ relation''. Lau {\it et al.}~\cite{lau2010} proposed a similar relation but with respect to the effective compactness (which is the inverse of the square root of the dimensionless moment of inertia, see Sec.~\ref{slow_rot}), and this relation was updated by~\cite{chirenti2015}. Chan {\it et al.}~\cite{chan2014} showed that the $f$-mode can also be related to the tidal deformability in an EOS-insensitive way. This relation is referred to as the ``$f$-Love relation'', and was further studied in~\cite{wen2019}, where implications from GW170817 were also explored.

        In BNS mergers, we can describe the ringdown phase of the GW signal by the QNMs of the remnant. The future detections of QNMs from long- or short-lived postmerger remnants, such as supramassive NSs (SMNSs) or hypermassive NSs (HMNSs), could carry important information about the EOS. In particular, numerical-relativity simulations have shown that the $f$-mode is the dominant mode in the GW spectrum of postmerger remnants~\cite{shibata2006,  hotokezaka2013_2, bernuzzi2014, lehner2016, depietri2018}. The future detection of GWs from such objects could establish constraints on the EOS through universal relations between the peak frequency and NS observables during the premerger phase of the GW signal~\cite{bauswein2012, bauswein2012_2,  bauswein2014, takami2015, bernuzzi2015, rezzolla2016, bauswein2017, vretinaris2020, blacker2020, lioutas2021}.

        Universal relations for NSs should be used with care when applying them to situations outside of their validity. In this work, we are interested in two scenarios:  protoneutron stars and  hypermassive neutron stars. We study the time evolution of the $I$-Love-$Q$ relations for PNSs using results from the state-of-the-art 3D CCSN simulations in Radice {\it et al.}~\cite{radice}, thus extending previous works~\cite{martinon, marques2017, raduta}. In particular, we study the $f$-Love relation by computing the $f$-mode for PNSs within full general relativity, improving on previous works that used Newtonian equations of motion~\cite{rodriguez2023, chaitanya2023}. We also study the evolution of the $f$-$C$ relation for HMNSs using results from the BNS simulations in Kastaun and Ohme~\cite{kastaun}. In contrast with recent works that relate the peak frequency of the HMNS with premerger quantities~\cite{rezzolla2016, breschi2019, lioutas2021}, we take a first step towards relating such frequencies with properties of the HMNS itself.
        
        The rest of this paper is organized as follows. In Sec.~\ref{sec:PNS}, we study properties and universal relations for PNSs (or at ``the birth'' of a NS). In Sec.~\ref{sec:HMNS}, we carry out similar studies but for HMNSs (or at ``the death'' of a NS).  We conclude in Sec.~\ref{disc}. Unless otherwise stated, we use $c = G = 1$ units.

    \section{``The birth'': protoneutron stars}
    
        \label{sec:PNS}

        We begin by studying properties and universal relations for PNSs, formed after CCSNe explosions. We use results from the 3D CCSN simulations described in Radice {\it et al.}~\cite{radice}, for which the SFHo EOS~\cite{steiner2013} was used, and eight different supernova progenitors were considered. In Table~\ref{tab_pns}, we provide some information about the models from the simulations. We consider 1D angle-averaged radial profiles for the pressure and total mass density from these simulations to construct a time-dependent effective barotropic EOS, or, simply, ``effective EOS'' (see Fig.~\ref{fig01}). We then obtain the structure of spherically symmetric PNSs by solving the Tolman-Oppenheimer-Volkoff (TOV) equations. We label these PNSs as ``TOV solutions'' (Sec.~\ref{tov_sol}). We finally use relativistic stellar perturbation theory to obtain their QNMs and $I$-Love-$Q$, using this effective EOS (Sec.~\ref{IIB}), and present the time evolution of the universal relations.

        \renewcommand{\thetable}{1}
        \begin{table}[t]
            \centering
            \begin{ruledtabular}
            \begin{tabular}{ccc}
                progenitor mass $[{\rm M}_{\odot}]$ & maximum time $[{\rm s}]$ & explode? \\
                \hline \hline
        	9 & 1.042 & yes \\
                \hline
                10 & 0.767 & yes \\
                \hline
                11 & 0.568 & yes \\
                \hline
                12 & 0.468 & yes \\
                \hline
                13 & 0.454 & no \\
        	\hline
        	19 & 0.871 & yes \\
                \hline
                25 & 0.616 & yes \\
                \hline
                60 & 0.398 & yes \\
            \end{tabular}
            \end{ruledtabular}
            \caption{Progenitor mass, maximum postbounce time, and explosion status for the 3D CCSN simulations in~\cite{radice}. We use the pressure and total mass density profiles from these models to construct time-dependent TOV solutions for PNSs (see Sec.~\ref{tov_sol}). The only model that does not explode within the simulation time is the one for the 13 M$_{\odot}$ progenitor (for more details, see~\cite{burrows2019, burrows2020}).}
            \label{tab_pns}
        \end{table}

        \subsection{TOV solutions}

            \label{tov_sol}
            
            PNSs right after the bounce have high temperature and  lepton-rich composition, as opposed to NSs. During the early postbounce phase, {\it i.e.} for a postbounce time $\lesssim 0.5$~s, the PNS mass is increasing as significant accretion is happening (and its rate depends on the density profile of the progenitor, see~\cite{burrows2019}), and the PNS radius is decreasing as neutrinos are slowly streaming out of the neutrinosphere. Nonetheless, since the hydrodynamical timescale in a PNS is $\sim 10^{-3}$~s (and decreases as the postbounce time increases), we can assume that hydrostatic equilibrium holds during the PNS evolution~\cite{burrows_lattimer, pons}.

            For a postbounce time $\gtrsim 0.5$~s, the PNS goes through the long-term Kelvin-Helmholtz phase, which can last tens of seconds. During this stage, the PNS deleptonizes and slowly cools down to become a ``cold'' NS~\cite{burrows_lattimer, pons}. The final NS is cold because its internal temperature ($\sim 10^{9}$~K) is much lower than the Fermi temperature of its composing nucleons ($\sim 10^{11}$~K). Thus, we refer to their EOS as cold ({\it viz.} zero-temperature) and barotropic, namely it is a function relating two thermodynamic variables, such as pressure and total mass density. For PNSs, however, this approximation is not valid as its internal temperature right after the bounce is higher ($\gtrsim 10^{11}$~K). Thus, PNSs are ``hot'' objects, and we refer to their EOS as hot and non-barotropic.

            The simulation data, provided by~\cite{radice}, are tables for various postbounce times $t_{\rm pb}$, with radial profiles for the quantities: pressure $p(t_{\rm pb}, \bar{r})$, baryonic mass density $\rho_{\rm B}(t_{\rm pb}, \bar{r})$, specific internal energy $e(t_{\rm pb}, \bar{r})$, adiabatic index for the fluid perturbations $\Gamma_{1}(t_{\rm pb}, \bar{r})$, and lapse function $\alpha(t_{\rm pb}, \bar{r})$; here, $\bar{r}$ is the isotropic radial coordinate, and we treat $t_{\rm pb}$ as a ``time coordinate'' since the profiles from the simulations correspond to specific postbounce times. We assume the EOS is effectively barotropic for each $t_{\rm pb}$ (similarly to~\cite{camelio1}). The total mass density $\rho(t_{\rm pb}, \bar{r})$ is given by $\rho = \rho_{\rm B}(1+e)$. Thus, we extract the relation $p(t_{\rm pb}, \rho)$ by eliminating $\bar{r}(t_{\rm pb}, \rho)$, which is obtained by inverting $\rho(t_{\rm pb}, \bar{r})$, from $p(t_{\rm pb}, \bar{r})$. This ``$p$~{\it vs.}~$\rho$'' relation is our time-dependent effective barotropic EOS, shown in Fig.~\ref{fig01} for the PNS generated by the 9~M$_{\odot}$ progenitor.

            \renewcommand{\thefigure}{1}
            \begin{figure}[t]
                \centering
                \includegraphics[width=0.49\textwidth]{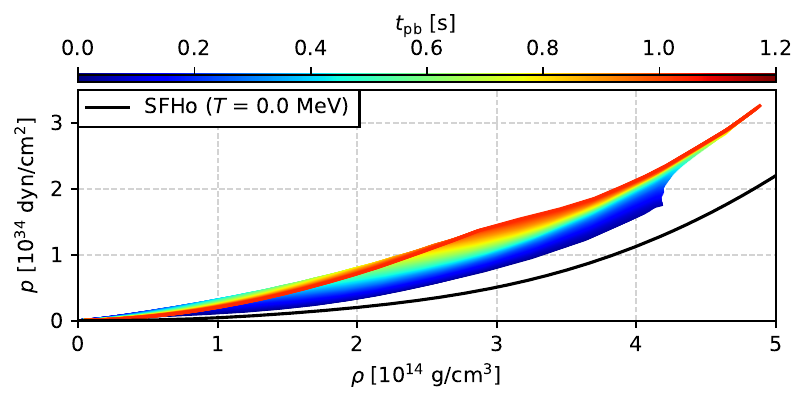}
                \caption{Relation between pressure~$p$ and total mass density~$\rho$ for various postbounce times~$t_\mathrm{pb}$ for the PNS generated by the 9 M$_{\odot}$ progenitor (see Table~\ref{tab_pns}), where the SFHo EOS~\cite{steiner2013} was used. For reference, we include the zero-temperature EOS.}
                \label{fig01}
            \end{figure}

            As a first step, we restrict ourselves to spherical symmetry and solve the TOV equations for the sequence of discrete $t_{\rm pb}$ of each progenitor mass. We characterize the PNSs by these equilibrium solutions or ``TOV solutions'', which are described by the radial profiles for the pressure $p(t_{\rm pb}, r)$, total mass density $\rho(t_{\rm pb}, r)$, gravitational mass $m(t_{\rm pb}, r)$, and gravitational potential $\Phi(t_{\rm pb}, r)$; here, $r$ is the Schwarzschild radial coordinate.

            For cold and spherically symmetric NSs, we usually define the circumferential radius $R$ through the condition $p(R) = 0$. However, we cannot impose this condition here because, for each $t_{\rm pb}$, there is no ``surface'' separating the ``interior'' of the PNS from its surroundings. The common choice in the literature is that the ``surface'' is defined by a baryonic mass density cutoff $\rho^{\rm srf}_{\rm B}$ normally taken as $\rho^{\rm srf}_{\rm B} = 10^{11}$~g/cm$^{3}$. In this work, we follow~\cite{morozova} and take $\rho^{\rm srf}_{\rm B} = 10^{10}$~g/cm$^{3}$, but we also generate results for $\rho^{\rm srf}_{\rm B} = 10^{11}$~g/cm$^{3}$ and $\rho^{\rm srf}_{\rm B} = 10^{12}$~g/cm$^{3}$, and present a comparison in Appendix~\ref{apA}. We can use the relation $p(t_{\rm pb}, \rho_{\rm B})$, which we construct similarly to $p(t_{\rm pb}, \rho)$, and determine the pressure cutoff $p^{\rm srf} = p(t_{\rm pb}, \rho^{\rm srf}_{\rm B})$. Then, the circumferential radius is given by $R(t_{\rm pb}) = r(t_{\rm pb}, p^{\rm srf})$, where we obtain $r(t_{\rm pb}, p)$ by inverting $p(t_{\rm pb}, r)$. The total gravitational mass is given by $M(t_{\rm pb}) = m(t_{\rm pb}, R)$. 

            In the left panel of Fig.~\ref{fig02}, we show the time evolution of $R$ and $M$ for the PNSs generated by the eight progenitors shown in Table~\ref{tab_pns}. We note the following features: (i) the circumferential radius is in a range $\approx$ 25 $-$ 125 km, and decreases with time; (ii) the gravitational mass is in a range $\approx$ 1.0 $-$ $1.8$ M$_{\odot}$, and increases with time.

            \renewcommand{\thefigure}{2}
            \begin{figure*}
                \centering
                \includegraphics[width=\textwidth]{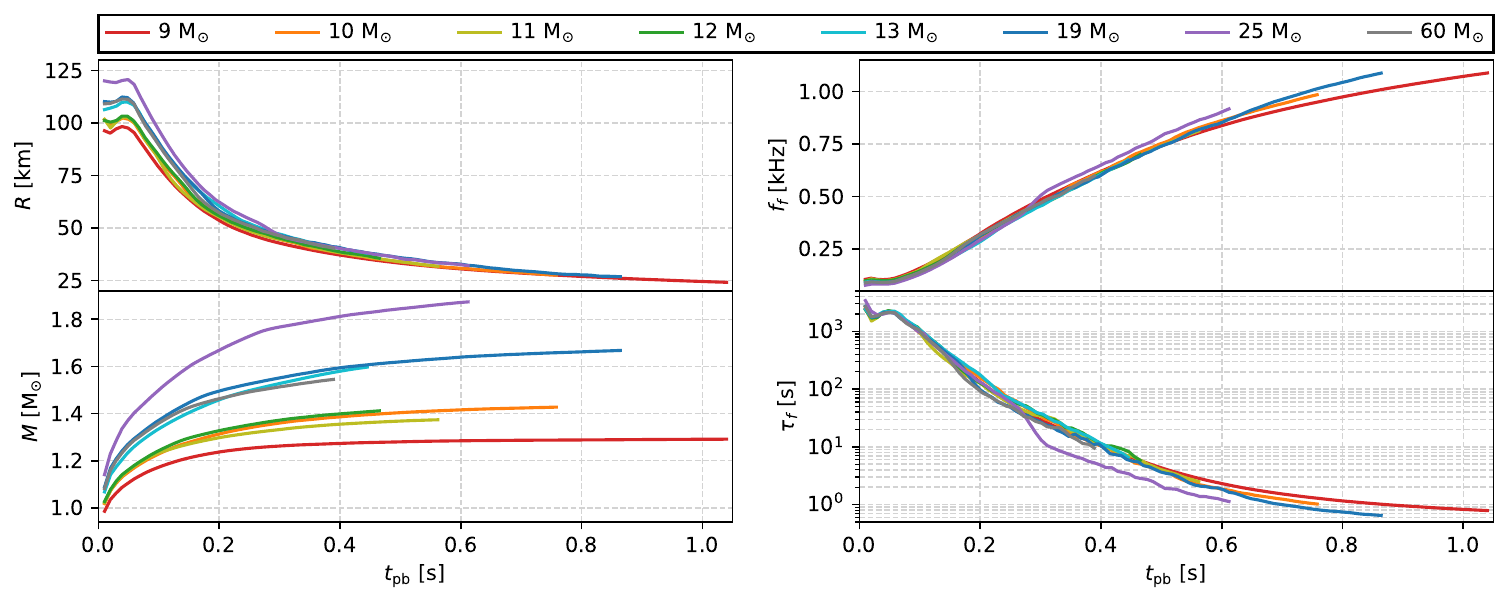}
                \caption{{\it Left panel}: Time evolution of the circumferential radius $R$ and the gravitational mass $M$ for PNSs generated by eight different progenitors (see Table~\ref{tab_pns}). The radius is determined by the density cutoff $\rho^{\rm srf}_{\rm B} = 10^{10}$~g/cm$^{3}$ (see main text). As expected, the mass increases and the radius decreases over time (asymptotically approaching cold NS values), owing to matter accretion and neutrino emission, respectively. {\it Right panel}: Time evolution of the oscillation frequency $f_{f}$ and the damping time $\tau_{f}$ of the $f$-mode for PNSs generated by eight different progenitors (see Table~\ref{tab_pns}). The frequencies are lower and the damping times are (much) longer, compared to the ones for cold NSs ($f^{\rm NS}_{f} \approx 1.5-3.0$ kHz and $\tau^{\rm NS}_{f} \approx 0.1-0.5$ s), due to the thermal effects on the effective EOS.}
                \label{fig02}
            \end{figure*}

            It is instructive to check if the radial profiles for our TOV solutions agree with the simulation data, and thus we show a comparison in Appendix~\ref{apB}. We obtained that the $p(t_{\rm pb},r)$ and $\rho(t_{\rm pb},r)$ profiles are similar, but the $m(t_{\rm pb},r)$ and $\Phi(t_{\rm pb},r)$ profiles are significantly different, by virtue of general-relativistic effects. These differences affect, for example, the radius and mass of our solutions, which are lower than the results obtained from the simulation by $\approx 6-12\%$ and $\approx 15-19\%$, for $t_{\rm pb} \gtrsim 0.2$~s (see Fig.~\ref{comp} in Appendix~\ref{apB}).
            
            The evolution of the radius and the mass has an impact on the evolution of other important PNS observables, such as frequencies of nonradial oscillations (as shown in Sec.~\ref{osc}). The study of radial oscillations is also important since it encodes the stability of the PNS against radial perturbations as it evolves. We investigate the radial stability of our TOV solutions in Appendix~\ref{apC}, and note that our PNSs are radially stable for all $t_{\rm pb}$, when entropy and composition gradients in the EOS are taken into account (see Fig.~\ref{fig03} in Appendix~\ref{apC}).

        \subsection{\textbf{\textit{I}}-Love-\textbf{\textit{Q}} and \textbf{\textit{f}}-Love for protoneutron stars}
    
            After obtaining the structure of spherically symmetric PNSs, we can perturb these background solutions, and compute PNS observables. We consider nonradial perturbations, small rotation, and small tidal deformation to obtain the PNS $f$-mode, moment of inertia, spin-induced quadrupole moment, and tidal deformability. Finally, we study the time evolution of universal relations involving these observables for PNSs, namely the $I$-Love-$Q$ and $f$-Love relations.

            \label{IIB}

            \subsubsection{Nonradial oscillations}

                \label{osc}
    
                Nonradial oscillations of PNSs can be excited, as a consequence of the turbulent nature of the collapse. These oscillations are coupled to the emission of GWs, and can be described by the QNMs of the PNSs. The GW signal from CCSNe has been extensively modeled, in particular, through 2D (axisymmetric,~\cite{finn1990, dimmelmeier2002, shibata2004, dimmelmeier2007, marek2009, murphy2009, kotake2009, muller2013, cerda2013, abdikamalov2014, yakunin2015, pan2018, morozova}) and 3D (\cite{mueller1997, rampp1998, fryer2004, shibata2005, ott2007, ott2011, ott2013, tamborra2013, kuroda2014, melson2015b, lentz2015, melson2015a, hayama2016, kuroda2016, roberts2016, yakunin2017, andresen2017, muller2017, kuroda2017, summa2018, ott2018, kuroda2018, hayama2018, oconnor2018, vartanyan2019, glas2019, andresen2019, powell2019, mezzacappa2020, vartanyan2022, nakamura2022, powell2022, mezzacappa2023, vartanyan2023}) simulations. Nevertheless, it was only recently that some works have proposed that the spectrogram of the GW signal from such simulations is dominated by the fundamental quadrupolar oscillation mode of the PNS \cite{fuller2015, morozova, torres2018, torres2019, radice, rodriguez2023}, making it one of the most relevant modes in the analysis of GW signals from CCSNe.

                In general, the complex frequencies of the QNMs are defined by $\omega \equiv 2\pi f+i\tau^{-1}$, where $f$ is the oscillation frequency and $\tau$ is the damping time. We can define the dimensionless $\omega$ as $\bar{\omega} \equiv M\omega$\footnote{This definition is not unique. We could have, {\it e.g.} $\tilde{\omega} \equiv M_{\rm B}\omega$ (see Sec.~\ref{sec:HMNS} and Appendix~\ref{apA}) or even $\hat{\omega} \equiv \omega/\sqrt{\rho_{\rm B,0}}$ (see the end of Sec.~\ref{osc}), where $M_{\rm B}$ is the total baryonic mass and $\rho_{\rm B,0}$ is the central baryonic mass density.}. The QNMs are divided into families~\cite{kokkotas, nollert}; for nonrotating and non-magnetized stars, we have: (i) $f$-, $p$-, and $g$-modes, which are fluid modes, and exist only for polar oscillations (even-parity perturbations); (ii) $w$-modes, which are spacetime modes, and exist for both polar and axial oscillations (the latter corresponding to odd-parity perturbations). Further, the QNMs can be classified by their restoring force, which is responsible for re-establishing the equilibrium on the perturbed star.

                \renewcommand{\thefigure}{3}
                \begin{figure*}
                    \centering
                    \includegraphics[width=\textwidth]{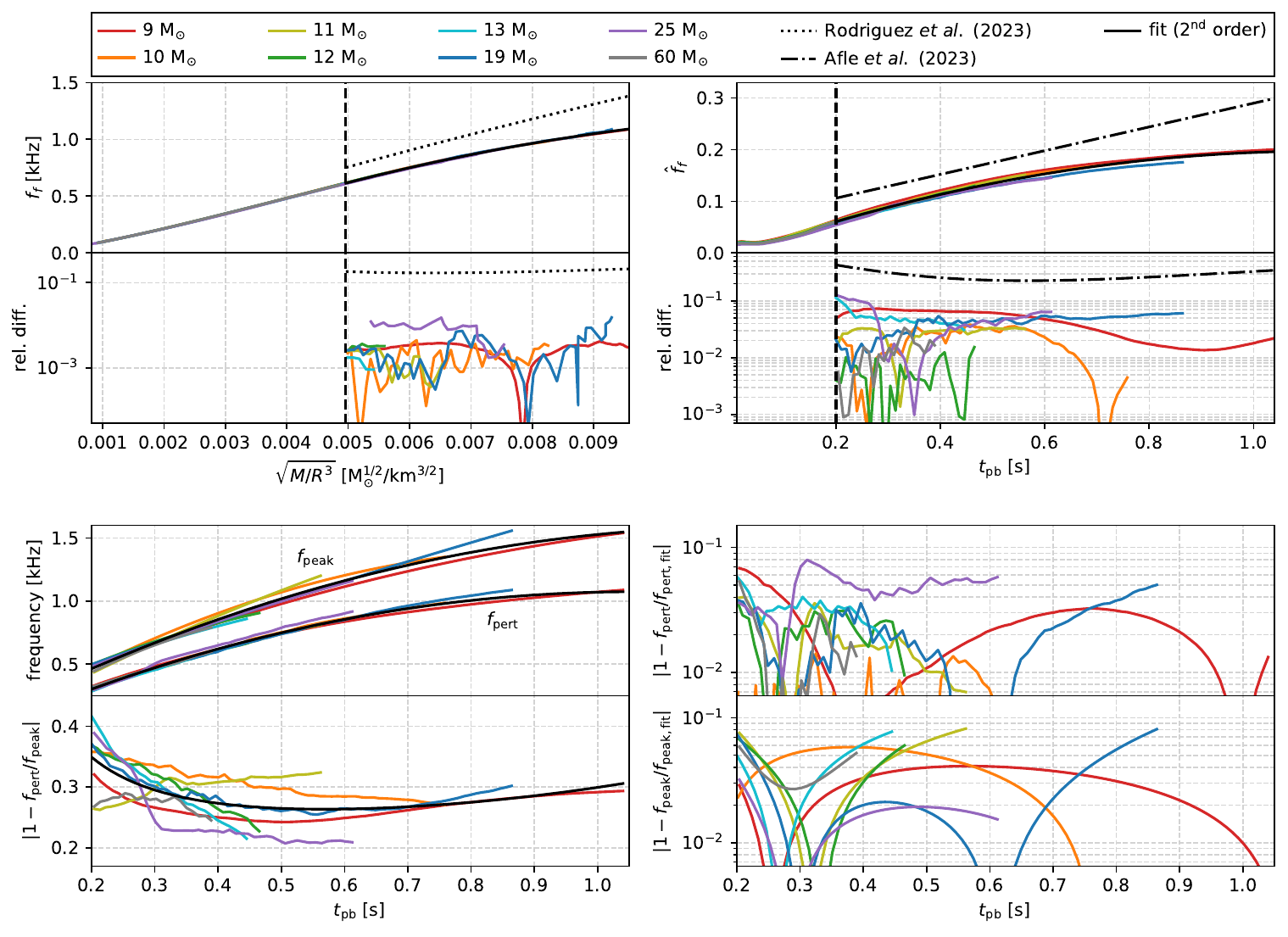}
                    \caption{{\it Upper left panel}: Relation between the $f$-mode frequency $f_{f}$ and the average density $\sqrt{M/R^{3}}$ for PNSs from eight progenitors. This relation is universal against the progenitor mass and the deviation is $|1-f_{f,{\rm fit}}/f_{f}| \lesssim 2\%$. The fitting coefficients are shown in Table~\ref{table_coeff_1}. For reference, we include the fit from Rodriguez {\it et al.}~\cite{rodriguez2023} (see main text), that we denote as $f_{f,{\rm R}}$, and we see that $|1-f_{f,{\rm fit}}/f_{f,{\rm R}}| \lesssim 21\%$. The vertical dashed line indicates the starting point ($\approx 0.005$~M$^{1/2}_{\odot}$/km$^{3/2}$) for the validity of the Cowling classification for the $f$-mode in~\cite{rodriguez2023}. {\it Upper right panel}: Relation between the dimensionless $f$-mode frequency ${\hat{f}}_{f}$ (see main text) and the postbounce time $t_{\rm pb}$ for PNSs from eight progenitors. This relation is not as tight as the relation between $f_{f}$ and $\sqrt{M/R^{3}}$ since the deviation is $|1-{\hat{f}}_{f,{\rm fit}}/{\hat{f}}_{f,{\rm num}}| \lesssim 12\%$. The fitting coefficients are shown in Table~\ref{table_coeff_1}. For reference, we include the fit from Afle {\it et al.}~\cite{chaitanya2023} (see main text), that we denote as ${\hat{f}}_{f,{\rm A}}$, and we see that $|1-{\hat{f}}_{f,{\rm fit}}/{\hat{f}}_{f,{\rm A}}| \lesssim 43\%$. The vertical dashed line indicates the starting point ($\approx 0.2$ s) for the $f$-mode in the GW spectrogram, as considered by~\cite{chaitanya2023}. {\it Lower left panel}: Time evolution of the dominant frequency extracted from the GW spectrograms, $f_{\rm peak}$ (see Appendix~\ref{apA}), and the $f$-mode frequency computed from perturbation theory, $f_{\rm pert}$ (see Sec.~\ref{osc}), for PNSs from eight progenitors. We fit the relations to a second-order fit and the relative difference between $f_{\rm pert}$ and $f_{\rm peak}$ is $\lesssim 43\%$, which agrees with the deviation in the relation between ${\hat{f}}_{f}$ and $t_{\rm pb}$ (not being the case for the relation between $f_{f}$ and $\sqrt{M/R^{3}}$, see main text). {\it Lower right panel}: Relative difference between $f_{\rm pert,fit}$ ($f_{\rm peak,fit}$) and $f_{\rm pert}$ ($f_{\rm peak}$). The progenitor-mass-variation is $\lesssim 9\%$.}
                    \label{comp_rod_afle}
                \end{figure*}

                We focus on the $f$-mode, or the fundamental mode, whose radial eigenfunction for the fluid perturbations has no nodes inside the star, following the Cowling classification \cite{cowling1941}. The $f$-mode of cold and nonrotating NSs has a frequency in the range $\approx 1.5 - 3.0$~kHz and a damping time in the range $\approx 0.1 - 0.5$~s. The frequencies of the $p$- and $g$-modes are, respectively, higher and lower than the $f$-mode frequency, while the damping times are higher.
                
                We investigate the time evolution of universal relations, involving QNMs, that are well established for cold and nonrotating NSs. Similarly to the $f$-mode, the relation between the $p_{1}$-mode frequency and the tidal deformability is also EOS-insensitive~\cite{sotani2021} (and the same has been found for the $g_{1}$-mode~\cite{kuan2022}). Nonetheless, since the GW spectrograms from the CCSN simulations described in~\cite{radice} are dominated by the $f$-mode frequency at late $t_{\rm pb}$, we do not show results for $p$- or $g$-modes. Finally, $w$-modes have not yet appeared in the GW spectrum of CCSN simulations, possibly indicating that, if excited, $w$-modes may have negligible amplitudes. However, given the relative simplicity of the eigenvalue problem for the axial $w$-modes (when compared to the $f$-mode), we report, in Appendix~\ref{apD}, results for the first curvature mode frequency and damping time (which are also universal against the tidal deformability~\cite{mena2019}).

                We use relativistic stellar perturbation theory to compute the $f$-mode complex frequency of our TOV solutions for the sequence of $t_{\rm pb}$ of each progenitor mass (see Sec.~\ref{tov_sol}). In particular, we follow the procedure of Lindblom and Detweiler~\cite{linddet}, {\it i.e.} we solve the perturbation equations in the form shown in~\cite{detweiler}, based on previous works~\cite{thorne, linddet}, and considering the corrections pointed out in~\cite{lu2011}. We impose regularity of the solutions at the center, and match them to the Zerilli function and its derivative at the surface. Then, we use a shooting method to find the complex frequency that gives us an outgoing wave solution for the Zerilli equation at infinity\footnote{In practice, we terminate the integration at $r=25/{\rm Re}(\omega)$, and match the solution with the asymptotic expansion given by~\cite{detweiler}.}. We take $\Gamma_{1} \approx \Gamma_{0}$ in the perturbation equations, where $\Gamma_{1}$ is the adiabatic index for the fluid perturbations and $\Gamma_{0} \equiv (\textrm{d}\ln{p}/\textrm{d}\ln{\rho_{\rm B}})_{\rm EOS}$ is the background adiabatic index\footnote{\label{note}Strictly speaking, $\Gamma_{0}$ cannot be referred as an adiabatic index, since the entropy is not kept constant inside the PNSs. For a discussion on adiabatic indices for PNSs, see Sec. 6 of~\cite{gondek}.} for the effective EOS. Thus, as $g$-modes are absent\footnote{The condition $\Gamma_{1}=\Gamma_{0}$ implies that the Schwarzschild discriminant is zero, and so are the frequencies of the $g$-modes~\cite{kokkotas}.}, we can use the Cowling classification for the modes, even at early $t_{\rm pb}$ (this is not the case in, {\it e.g.}~\cite{rodriguez2023}).
                
                In the right panel of Fig.~\ref{fig02}, we show the results for the oscillation frequency $f(t_{\rm pb})$ and damping time $\tau(t_{\rm pb})$ for the PNSs generated by the eight progenitors shown in Table~\ref{tab_pns}. The frequencies of the $f$-mode for hot and young NSs have already been studied in, {\it e.g.}~\cite{ferrari2003, burgio2011, camelio1, thapa2023}. In general, thermal effects are expected to lower the frequency of the oscillation mode, since thermal pressure increases the radius of the stars, and thus decreases their compactness. In the right panel of Fig.~\ref{fig02}, we see that the frequencies for the PNSs are in a range $\approx 0.1 - 1.0$~kHz, which is indeed lower than the typical NS frequencies. We also see that the damping times are in a range $\approx 10 - 10^{3}$~s, much longer than those for cold NSs. These lower frequencies and larger damping times make the future detection of this mode by third-generation ground-based GW detectors more promising if they are excited with enough energy. We also performed the calculation of the $f$-mode considering the adiabatic index $\Gamma_{1}$ from the simulations when solving the perturbation equations, and we verified that the frequencies and damping times agree with the ones obtained by considering $\Gamma_{1} \approx \Gamma_{0}$ for late $t_{\rm pb}$ (see Fig.~\ref{fmode_g1} in Appendix~\ref{apC}).

                Once we have computed the $f$-mode for different time slices and progenitor masses, we can investigate universal relations for PNSs involving this mode. Following Rodriguez {\it et al.}~\cite{rodriguez2023}, we study the relation between the $f$-mode frequency $f_{f}$ and the average density $\sqrt{M/R^{3}}$ of the PNSs. We show this result in the upper left panel of Fig.~\ref{comp_rod_afle}, where we can see that the relation is universal against the progenitor mass. We also present the fit proposed by~\cite{rodriguez2023}, which also used the results from the CCSN simulations described in~\cite{radice}. The $f$-mode frequencies in their fit are obtained in the Cowling approximation and follow a different classification (the classification based on modal properties, see~\cite{rodriguez2023} for more details), that matches the Cowling classification at late $t_{\rm pb}$ ($t_{\rm pb} \gtrsim 0.4$ s), {\it i.e.} high $\sqrt{M/R^{3}}$ ($\gtrsim 0.005$~M$^{1/2}_{\odot}$/km$^{3/2}$). Thus, we present a comparison between the relations for the same range in $\sqrt{M/R^{3}}$. We see that: (i) the relative difference $|1-f_{f}/f_{f,{\rm fit}}|$, between our fit $f_{f,{\rm fit}}$ and our data $f_{f}$, is $\lesssim 2\%$; (ii) the relative difference $|1-f_{f,{\rm fit}}/f_{f,{\rm R}}|$, between our fit $f_{f,{\rm fit}}$ and their fit $f_{f,{\rm R}}$, is $\lesssim 21\%$. The fitting coefficients are shown in Table~\ref{table_coeff_1}. 

                \renewcommand{\thetable}{2}
                \begin{table*}
                    \centering
                    \begin{ruledtabular}
                        \begin{tabular}{ccccc}
                            $y$ & $\sqrt{M/R^{3}}$ [M$^{1/2}_{\odot}$/km$^{3/2}$] & $t_{\rm pb}$ [s] & $t_{\rm pb}$ [s] & $t_{\rm pb}$ [s] \\
                            \hline
                            $x$ & $f_{f}$ [kHz] & ${\hat{f}}_{f}$ & $f_{\rm pert}$ [kHz] & $f_{\rm peak}$ [kHz] \\
                            \hline \hline
                            $c_{0}$ & $-$2.632$\times10^{-1}$ & $-$4.875$\times10^{-3}$ & $-$1.013$\times10^{-1}$ & $-$9.204$\times10^{-3}$ \\
                            \hline
                            $c_{1}$ & 2.147$\times10^{2}$ & 3.587$\times10^{-1}$ & 2.229 & 2.574 \\
                            \hline
                            $c_{2}$ & $-$7.655$\times10^{3}$ & $-$1.593$\times10^{-1}$ & $-$1.057 & $-$1.036 \\
                            \hline \hline
                            $\Delta y^{\rm rms}$ & 4.288$\times10^{-3}$ & 4.063$\times10^{-2}$ & 2.735$\times10^{-2}$ & 3.645$\times10^{-2}$ \\
                           \hline
                            $\Delta y^{\rm max}$ & 1.519$\times10^{-2}$ & 1.231$\times10^{-1}$ & 7.956$\times10^{-2}$ & 8.181$\times10^{-2}$ \\
                        \end{tabular}
                    \end{ruledtabular}
                    \caption{Fitting coefficients for the relations: $f_{f}$ {\it vs.} $\sqrt{M/R^{3}}$ (upper left panel of Fig.~\ref{comp_rod_afle}), ${\hat{f}}_{f}$ {\it vs.} $t_{\rm pb}$ (upper right panel of Fig.~\ref{comp_rod_afle}), and $f_{\rm pert}$ {\it vs.} $t_{\rm pb}$ as well as $f_{\rm peak}$ {\it vs.} $t_{\rm pb}$ (lower left panel of Fig.~\ref{comp_rod_afle}). The functional form of the fit is $y_{\rm fit}=\sum_{i=0}^{n}c_{i}x^{i}$, where $n = 2$. The relative difference is $\Delta y=|1-y_{\rm fit}/y|$, and we show the root mean square $\Delta y^{\rm rms}$ and the maximum $\Delta y^{\rm max}$.}
                    \label{table_coeff_1}
                \end{table*}

                Following Afle {\it et al.}~\cite{chaitanya2023}, we can also investigate the relation between the $f$-mode frequency, normalized by the central baryonic mass density $\rho_{\rm B,0}$, and the postbounce time $t_{\rm pb}$. We define ${\hat{\omega}}_{f} \equiv \omega_{f}/\sqrt{\rho_{\rm B,0}}$, and show the relation between ${\hat{f}}_{f}\equiv{\rm Re}({\hat{\omega}}_{f})/2\pi$ and $t_{\rm pb}$ in the upper right panel of Fig.~\ref{comp_rod_afle}, where we note that the relation is universal against $t_{\rm pb}$. We also present the fit proposed by~\cite{chaitanya2023}, that also used the results from the CCSN simulations described in~\cite{radice}. The $f$-mode frequencies in their fit were obtained from the GW spectrograms. The initial time for the $f$-mode (namely, the time such that $f$-mode starts contributing to the GW signal) in their work was set to 0.2 s, thus we show a comparison between the relations for the same initial time. We see that (i) the relative difference $|1-{\hat{f}}_{f}/{\hat{f}}_{f,{\rm fit}}|$, between our fit ${\hat{f}}_{f,{\rm fit}}$ and our data ${\hat{f}}_{f}$, is $\lesssim 12\%$; (ii) the relative difference $|1-{\hat{f}}_{f,{\rm fit}}/{\hat{f}}_{f,{\rm A}}|$, between our fit ${\hat{f}}_{f,{\rm fit}}$ and their fit ${\hat{f}}_{f,{\rm A}}$, is $\lesssim 43\%$. The fitting coefficients are shown in Table~\ref{table_coeff_1}.

                When comparing our results with Rodriguez {\it et al.}~\cite{rodriguez2023}, we note that our frequencies, for the same average densities, are lower. Similarly, when comparing our results with Afle {\it et al.}~\cite{chaitanya2023}, we see that our normalized frequencies, for the same postbounce times, are lower. In the latter case, since we are normalizing the $f$-mode frequency by the central baryonic mass density $\rho_{\rm B,0}$, which we take from the simulation data, this comparison is equivalent to a direct comparison between the frequencies. In the lower panels of Fig.~\ref{comp_rod_afle}, we show a comparison between the frequencies from the GW spectrograms (see Appendix~\ref{apA} for more details), that we denote by $f_{\rm peak}$, and the frequencies from perturbation theory (see Sec.~\ref{osc}), that we denote by $f_{\rm pert}$. We obtain that (i) the relative difference $|1-{f_{\rm pert}}/{f_{\rm peak}}|$, between $f_{\rm pert}$ and $f_{\rm peak}$, is $\lesssim 43\%$ (lower left panel); (ii) the progenitor-mass-variation in the relations, given by the relative differences $|1-{f_{\rm pert}}/{f_{\rm pert,fit}}|$ and $|1-{f_{\rm peak}}/{f_{\rm peak,fit}}|$, is $\lesssim 9\%$ (lower right panel).

                Therefore, we obtain that the source of the difference between our results and Afle {\it et al.}~\cite{chaitanya2023} ($\lesssim 43\%$) is the modeling of the PNSs. We solve the general-relativistic equations to obtain background and perturbed PNSs, as opposed to~\cite{radice}, that adopted the conformally flat approximation to obtain the simulated PNSs, and the quadrupole formula to obtain their GW signal\footnote{The dominant frequencies from the GW spectrograms in~\cite{chaitanya2023} match the $f$-mode frequencies obtained with the perturbation analysis, using Newtonian equations of motion, shown in~\cite{morozova}.} (see Appendix~\ref{apB} for more details).
                
                We attribute the lower difference between our results and Rodriguez {\it et al.}~\cite{rodriguez2023} ($\lesssim 21\%$) to the Cowling approximation adopted in the  $f$-mode calculation\footnote{In~\cite{rodriguez2023}, the formalism used for the calculation of the $f$-mode was also the one presented in~\cite{morozova}, and the Cowling approximation reduces the frequency, as shown in Fig. 5 of~\cite{rodriguez2023}}. The differences in the modeling of the PNSs do not affect their average density significantly (with a relative difference $\lesssim 10\%$, see Fig.~\ref{comp} in Appendix~\ref{apB}).

            \subsubsection{Slow rotation}

                \label{slow_rot}

                \renewcommand{\thefigure}{4}
                \begin{figure*}
                    \centering
                    \includegraphics[width=\textwidth]{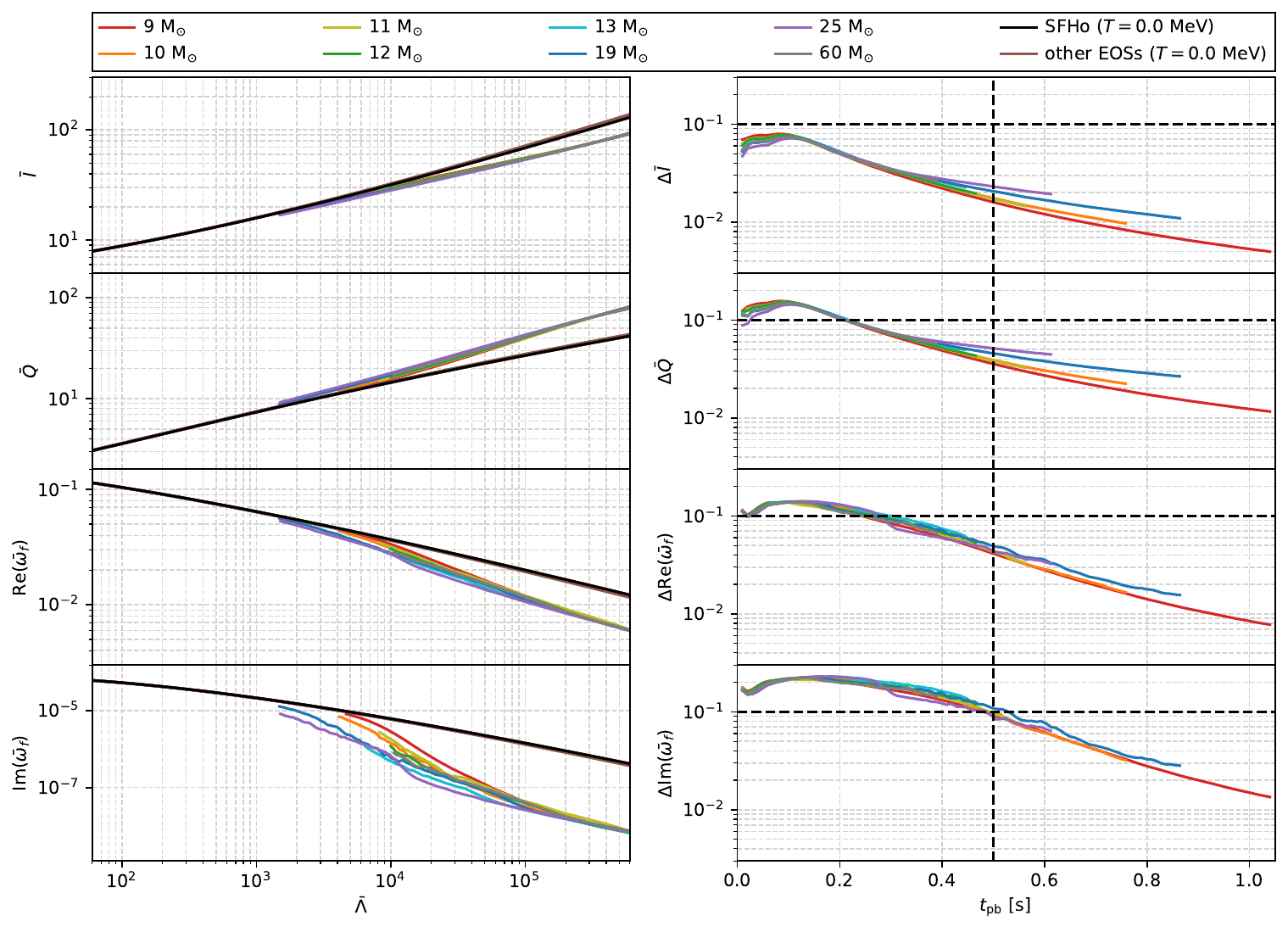}
                    \caption{{\it Left panel}: Time evolution of $I$-Love-$Q$ and $f$-Love relations for PNSs generated by eight different progenitors (see Table~\ref{tab_pns}). We consider $\bar{\Lambda}$ as the independent variable and show its correlation with $\bar{I}$ and $\bar{Q}$, and with Re($\bar{\omega}_{f}$) and Im($\bar{\omega}_{f}$). For reference, we include the relations for cold NSs, considering ten zero-temperature EOSs (see main text). We note that, as the postbounce time $t_{\rm pb}$ increases (from high $\bar{\Lambda}$ to low $\bar{\Lambda}$), the PNS relations approach the NS relations. {\it Right panel}: Time evolution of the relative difference $\Delta y=|1-y_{\rm PNS}/y_{\rm NS}|$ between PNS observables $y_{\rm PNS}$ and the NS observables $y_{\rm NS}$, where $y = y(x)$, for $x = \bar{\Lambda}$ and $y \in \{\bar{I},\textrm{ }\bar{Q},\textrm{ }{\rm Re}({\bar{\omega}}_{f}),\textrm{ }{\rm Im}({\bar{\omega}}_{f})\}$. We note that $\Delta y$ decreases as $t_{\rm pb}$ increases. In particular, $\Delta y\lesssim 10\%$ for $t_{\rm pb} \gtrsim 0.5$~s (see horizontal and vertical dashed lines).}
                    \label{ilq_err}
                \end{figure*}

                PNSs could be fast- and differentially-rotating right after the bounce, by virtue of angular momentum conservation and the nonaxisymmetric nature of the collapse~(see, {\it e.g.}~\cite{ott2006}). However, during the long-term Kelvin-Helmholtz phase, the angular momentum loss through neutrino emission makes their angular velocity decrease~\cite{mikaelian1977, kazanas1977, epstein1978, henriksen1978}. On top of that, viscosity mechanisms such as the magnetorotational instability~\cite{wheeler2015, mosta2015}, remove their differential rotation in a comparably short timescale. It is then reasonable that we use the slow and uniform rotation approximation to model rotating PNSs. We add slow and uniform rotation to our TOV solutions\footnote{In the CCSN simulations described in~\cite{radice}, only nonrotating progenitors were considered.}, and compute their moment of inertia $I(t_{\rm pb})$ and spin-induced quadrupole moment $Q(t_{\rm pb})$, for the eight progenitors shown in Table~\ref{tab_pns}, by implementing the Hartle-Thorne formalism \cite{hartle, hartle2}. We define the dimensionless $I$ as $\bar{I} \equiv I/M^{3}$, and the dimensionless $Q$ as $\bar{Q} \equiv Q/(\chi^{2}M^{3})$; here, $\chi \equiv J/M^{2}$ is the dimensionless spin, where $J$ is the spin angular momentum.

            \subsubsection{Small tidal deformation}
                
                The evolutionary paths of main-sequence star binaries can have PNSs in binary systems with main sequence stars, white dwarfs, or NSs~\cite{stairs}. Thus, in principle, PNSs could be tidally deformed. However, the separation distance between the stars in the binary would be too large and make tidal effects almost negligible, even in a common envelope scenario\footnote{The radius of the progenitor star is $\sim 10^{6}$ km and the radius of the core right before collapse is $\sim 10^{3}$ km.}. Still, we add small tidal deformation to our TOV solutions and compute their tidal deformability $\Lambda(t_{\rm pb})$, for the eight progenitors shown in Table~\ref{tab_pns}, following the procedure of Hinderer~\cite{tanja0} and Hinderer {\it et al.}~\cite{tanja}. We define the dimensionless $\Lambda$ as $\bar{\Lambda} \equiv \Lambda/M^{5}$, and we treat it as the varying parameter in the universal relations (as we show in Sec.~\ref{ilq}).

            \subsubsection{Time evolution of {\it I}-Love-{\it Q} and {\it f}-Love}

                \label{ilq}

                Once we have computed all of the relevant PNS observables, namely: the dimensionless $f$-mode complex frequency ${\bar{\omega}}_{f}$, moment of inertia $\bar{I}$, spin-induced quadrupole moment $\bar{Q}$, and tidal deformability $\bar{\Lambda}$, we can explicitly verify whether well-established universal relations for cold NSs are satisfied for PNSs. Here, we focus on the $I$-Love-$Q$ and $f$-Love relations, which have been shown to be the tightest universal relations between NS observables (in Appendix \ref{apD}, we show results for the $w$-Love relation).

                In the left panel of Fig.~\ref{ilq_err}, we take $\bar{\Lambda}$ as the independent variable and investigate its correlation with $\bar{I}$ and $\bar{Q}$, and with Re(${\bar{\omega}}_{f}$) and Im(${\bar{\omega}}_{f}$). For comparison, we also show the same relations for ten zero-temperature EOSs, including SFHo. We consider EOS models for npe$\mu$ nuclear matter (AP3~\cite{akmal_AP3}, FPS~\cite{friedman_FPS}, MPA1~\cite{muther_MPA1}, MS1~\cite{muller_MS1}, SFHo~\cite{steiner2013}, SLy4~\cite{douchin_SLy}, and WFF1~\cite{wiringa_WFF1}), as well as models that include hyperons (GNH3~\cite{glendenning_GNH3} and H4~\cite{lackey_H4}) and color-flavored-locked quark matter (ALF2~\cite{alford_ALF2}). We can see that the $I$-Love-$Q$ and $f$-Love relations for PNSs are different from those for cold NSs, and the former have a slight dependence on the progenitor mass. However, as the postbounce time $t_{\rm pb}$ increases (from high $\bar{\Lambda}$ to low $\bar{\Lambda}$), the PNS relations approach the cold NS relations. We found that the relations are approximately recovered in $\approx 1$ s, confirming the result of Martinon {\it et al.}~\cite{martinon} for the $I$-Love-$Q$ relations. There are two caveats: (i) these results are for the SFHo EOS only, and, unfortunately, simulations for other EOSs are not available; (ii) we have considered that the PNS density cutoff is $\rho^{\rm srf}_{\rm B} = 10^{10}$ g/cm$^{3}$, but in Appendix~\ref{apB}, we show results for $\rho^{\rm srf}_{\rm B} = 10^{11}$~g/cm$^{3}$ and $\rho^{\rm srf}_{\rm B} = 10^{12}$~g/cm$^{3}$, and we obtain that the universality is recovered earlier, as $\rho^{\rm srf}_{\rm B}$ increases.
                
                We can investigate the time evolution of the relative difference $\Delta y=|1-y_{\rm PNS}/y_{\rm NS}|$ between the PNS observables $y_{\rm PNS}$ and the NS observables $y_{\rm NS}$, where $y = y(x)$, for $x = \bar{\Lambda}$ and $y \in \{\bar{I},\textrm{ }\bar{Q},\textrm{ }{\rm Re}({\bar{\omega}}_{f}),\textrm{ }{\rm Im}({\bar{\omega}}_{f})\}$. We compute $\Delta y$ considering the NS observables for the SFHo EOS (since this was the EOS used in the simulations in~\cite{radice}). We show this result in the right panel of Fig.~\ref{ilq_err}, for the eight progenitors shown in Table~\ref{tab_pns}, where we can see that $\Delta y$ decreases as the postbounce time $t_{\rm pb}$ increases. These results show that the universal relations for cold NSs that we are considering here are approximately valid for PNSs at late postbounce times, {\it i.e.} $\Delta y \approx 1\% - 10\%$ for $t_{\rm pb} \gtrsim 0.5$~s. Thus, we extend the results in~\cite{martinon, marques2017, raduta} by using state-of-the-art 3D CCSN simulations, and show that the $f$-Love relation is also recovered within the time that the $I$-Love-$Q$ relations are recovered ($\approx$ 1 s), besides giving an estimate for the relative error of these relations for earlier times, which could be useful in future observations (see discussion in Sec.~\ref{disc}).

    \section{``The death'': hypermassive neutron stars}

        \label{sec:HMNS}

        \renewcommand{\thetable}{3}
        \begin{table}
            \centering
            \begin{ruledtabular}
                \begin{tabular}{ccc}
                    total baryonic mass $[{\rm M}_{\odot}]$ & mass ratio & maximum time $[{\rm ms}]$ \\
                    \hline \hline
                    3.008 & 0.9 & $\approx$ 9 \\
                    \hline
                    3.001 & 1.0 & $\approx$ 12 \\
                \end{tabular}
            \end{ruledtabular}
            \caption{Total baryonic mass, mass ratio, and maximum postmerger time (time for the formation of the apparent horizon) for the numerical-relativity BNS simulations in \cite{kastaun}. In both cases, the chirp mass is 1.187 M$_{\odot}$, consistent with the measured value for GW170817 ($1.186 \pm 0.001$ M$_{\odot}$)~\cite{ligo}.}
            \label{tab_hmns}
        \end{table}

        We next study the properties of HMNSs, formed after BNS mergers. We use results from numerical-relativity simulations of BNS mergers, from Kastaun and Ohme \cite{kastaun}, for two mass ratios: 0.9 and 1.0, where the SFHo EOS \cite{steiner2013} was used. In Table~\ref{tab_hmns}, we give some details about the models from the simulations. Motivated by the discussion of bulk properties of HMNSs by~\cite{kastaun} (see more in Sec.~\ref{coretov}), we obtain a new universal relation between bulk quantities of cold and nonrotating NSs, the ``bulk $f$-$C$ relation'', and study the evolution of HMNSs on the plane of this new universal relation.

        \subsection{TOV core equivalents}

            \label{coretov}

            When two NSs merge, there are four possible outcomes for the postmerger remnant\footnote{The outcome depends primarily on the masses of the stars in the binary and on their EOS.}~\cite{sarin}: (i) formation of a stable NS; (ii) formation of a SMNS, which is supported against collapse by uniform rotation and either collapses to a black hole on timescales $\sim$ 1 s $-$ 1 yr or forms a stable NS; (iii) formation of a HMNS, which is supported against collapse by differential rotation and either collapses to a black hole on timescales $\sim$ 10 ms $-$ 100 ms or evolves to a SMNS; (iv) prompt collapse to a black hole.

            \renewcommand{\thefigure}{5}
            \begin{figure}
                \centering
                \includegraphics[width=0.49\textwidth]{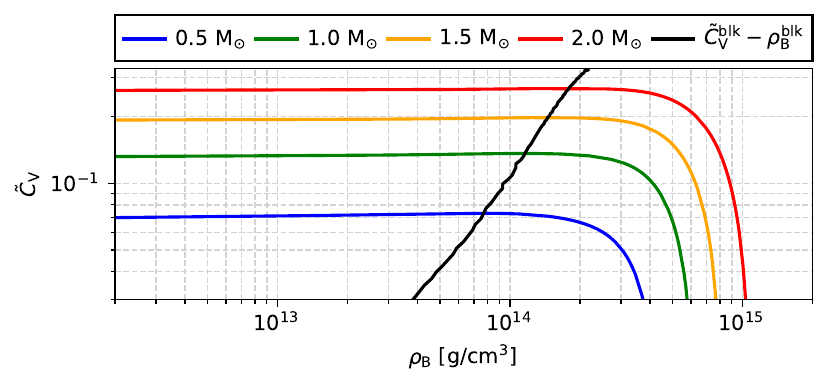}
                \caption{Relation between the compactness ${\tilde{C}}_{\rm V} = M_{\rm B}/R_{\rm V}$ and baryonic mass density $\rho_{\rm B}$ for 0.5~M$_{\odot}$, 1.0~M$_{\odot}$, 1.5~M$_{\odot}$, and 2.0~M$_{\odot}$ NSs described by the zero-temperature SFHo EOS. We also show the relation between the compactness of the bulk ${\tilde{C}}^{\rm blk}_{\rm V} = M^{\rm blk}_{\rm B}/R^{\rm blk}_{\rm V}$ and the surface density of the bulk $\rho^{\rm blk}_{\rm B}$ for the same EOS.}
                \label{bulk_plot}
            \end{figure}

            \renewcommand{\thefigure}{6}
            \begin{figure*}
                \centering
                \includegraphics[width=\textwidth]{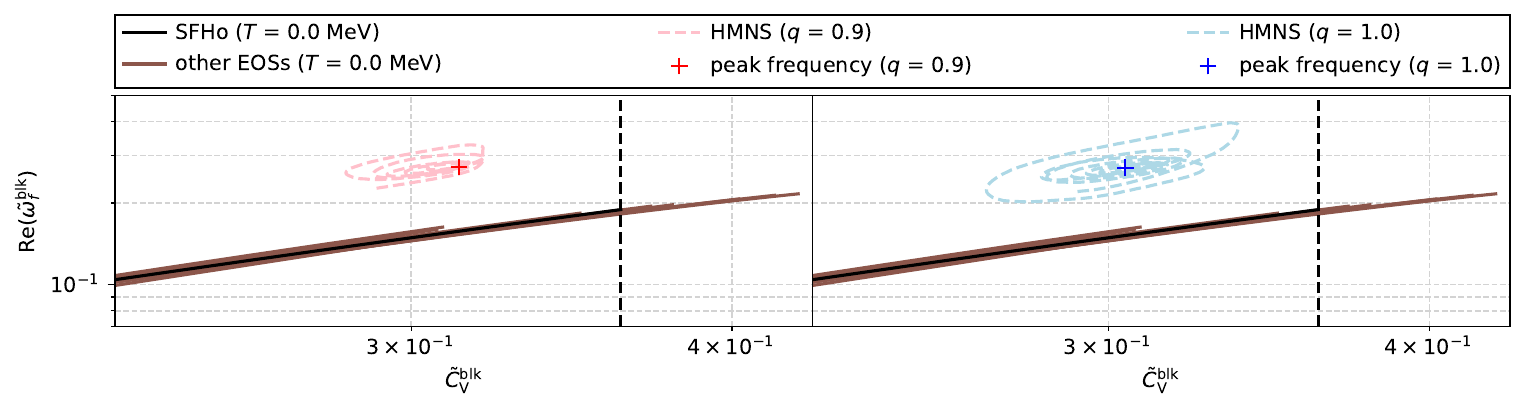}
                \caption{Relation between the dimensionless $f$-mode frequency of the bulk ${\rm Re}({\tilde{\omega}}^{\rm blk}_{f}) = {\rm Re}(M^{\rm blk}_{\rm B}\omega^{\rm blk}_{f})$ and compactness of the bulk ${\tilde{C}}^{\rm blk}_{\rm V} = M^{\rm blk}_{\rm B}/R^{\rm blk}_{\rm V}$, considering ten zero-temperature EOSs (see main text). The ``spiraling'' dashed curves show the instantaneous GW frequency of the simulated HMNSs on the ${\rm Re}({\tilde{\omega}}^{\rm blk}_{f})-{\tilde{C}}^{\rm blk}_{\rm V}$ plane and the ``plus'' markers are the normalized peak frequencies, for two mass ratios $q$ (see Table~\ref{tab_hmns}). The relative difference $|1-y_{\rm sim}/y_{\rm SFHo}|$ between the normalized peak frequency from the simulation $y_{\rm sim}({\tilde{C}}_{\rm V}^{\rm blk})$ and the frequency for the cold and nonrotating NS described by the SFHo EOS $y_{\rm SFHo}({\tilde{C}}_{\rm V}^{\rm blk})$, for the same compactness ${\tilde{C}}_{\rm V}^{\rm blk}$, is $\approx$ 73$\%$ for $q = 0.9$ and $\approx$ 78$\%$ for $q = 1.0$, where $y = {\rm Re}({\tilde{\omega}}_{f}^{\rm blk})$. The vertical dashed line indicates the maximum compactness of the bulk for the SFHo EOS, beyond which the simulated HMNS collapses to a black hole.}
                \label{hmns_plot}
            \end{figure*}

            HMNSs are rapidly- and differentially-rotating stars, described by a hot EOS \cite{baumgarte2000, sarin}. We cannot study these objects' structure using the approximations we used for PNSs. To obtain, {\it e.g.} the total baryonic mass of a HMNS at a postmerger time $t_{\rm pm}$, we need the baryonic mass density $\rho_{\rm B}$ as a function of the spatial coordinates at $t_{\rm pm}$. This function cannot be obtained trivially from BNS simulations, since HMNSs are highly nonaxisymmetric, which brings ambiguity in specifying the spatial dependence of $\rho_{\rm B}$.
    
            To overcome this, Kastaun {\it et al.}~\cite{kastaun2} proposed a measure for the description of postmerger remnants that does not depend on spatial coordinates, although it depends on the foliation of the spacetime. For a given time slice $t_{\rm pm}$, we consider a surface of constant $\rho_{\rm B}$ measured in the fluid rest frame. Thus, for each isodensity surface, we can measure the proper three-volume $V$, thus the volumetric radius $R_{\rm V} = (3V/4\pi)^{1/3}$, and the enclosed baryonic mass $M_{\rm B}$. Then, we have a $M_{\rm B}$ {\it vs.} $R_{\rm V}$ relation for each $t_{\rm pm}$. We can use this relation as a replacement for the $\rho_{\rm B}$ function, {\it i.e.} we can determine the total baryonic mass by specifying the total radius. We can also define a new compactness in terms of $M_{\rm B}$ and $R_{\rm V}$, given by ${\tilde{C}}_{\rm V} \equiv M_{\rm B}/R_{\rm V}$. Kastaun {\it et al.} \cite{kastaun2} defined the ``bulk'' as the region enclosed by the isodensity surface $\rho^{\rm blk}_{\rm B}$ that gives the maximum compactness ${\tilde{C}}^{\rm blk}_{\rm V}$ at each $t_{\rm pm}$. Similarly, the corresponding volumetric radius and baryonic mass of the bulk are given by $R^{\rm blk}_{\rm V}$ and $M^{\rm blk}_{\rm B} = M_{\rm B}(R^{\rm blk}_{\rm V})$, respectively.
    
            Even though this measure was defined for nonaxisymmetric postmerger remnants (and was further used in~\cite{ciolfi2017, kastaun2017, endrizzi2018, kastaun}), we can use it for cold and nonrotating NSs. It has been shown that the $M_{\rm B}$ {\it vs.} $R_{\rm V}$ profiles of cores of postmerger remnants are similar to those of NSs. The TOV solutions that best approximate the cores of postmerger remnants are called ``TOV core equivalents''. Further, it has been conjectured that if a postmerger remnant does not admit a TOV core equivalent, it promptly collapses to a black hole~\cite{ciolfi2017}. Moreover, if it admits a TOV core equivalent, the remnant collapses to a black hole when the $M_{\rm B}$ {\it vs.} $R_{\rm V}$ profile of this remnant matches that of the maximum-mass TOV core equivalent. If proven, this conjecture would allow us to accurately model the cores of long- or short-lived postmerger remnants with TOV core equivalents, with no concerns about radial stability (before the collapse to a black hole). In particular, the HMNSs in~\cite{kastaun} that we consider in this work admit TOV core equivalents.
    
            To illustrate the definition of the bulk (which delimits the TOV core equivalent), we show, in Fig.~\ref{bulk_plot}, the relation between ${\tilde{C}}_{\rm V}$ and $\rho_{\rm B}$ for four NS masses, as well as the ${\tilde{C}}^{\rm blk}_{\rm V}$ {\it vs.} $\rho^{\rm blk}_{\rm B}$ relation, using the zero-temperature SFHo EOS. We note that ${\tilde{C}}_{\rm V}$ is almost insensitive to the low-density parts of the star (note the slope of the curves for $\rho^{\rm srf}_{\rm B}\lesssim\rho^{\rm blk}_{\rm B}$ in Fig.~\ref{bulk_plot}). As pointed out in~\cite{kastaun2}, ${\tilde{C}}^{\rm blk}_{\rm V}$ could be a good candidate for a universal-relation parameter with respect to other NS observables, such as the QNM frequencies. Indeed, we can calculate the QNM frequencies of the bulk $\omega^{\rm blk}$, and relate the real and imaginary parts of their dimensionless counterparts, ${\tilde{\omega}}^{\rm blk} \equiv M^{\rm blk}_{\rm B}\omega^{\rm blk}$, with ${\tilde{C}}^{\rm blk}_{\rm V}$. In Fig.~\ref{hmns_plot}, we present this relation for the $f$-mode, considering ten zero-temperature EOSs, including SFHo (similarly to Fig.~\ref{ilq_err}). The EOS-variation is $\lesssim 3\%$ for a fourth-order fit (see Table~\ref{table_coeff_2} in Appendix~\ref{apE}).
            
            Note that the $f$-$C$ relation shown in Fig.~\ref{hmns_plot} involves the normalized $f$-mode frequency of the bulk and the compactness of the bulk. These quantities are defined in terms of the baryonic mass of the bulk $M^{\rm blk}_{\rm B}$ and the volumetric radius of the bulk $R^{\rm blk}_{\rm V}$. This relation is slightly different from the one widely used in previous literature ({\it e.g.}~\cite{tsui2005}), since their dimensionless frequency and compactness are defined in terms of the gravitational mass and circumferential radius of the NSs. In Appendix~\ref{apE}, we compare these different relations involving the $f$-mode and the compactness for different definitions.
            
            We obtain that the different relations are similar when the compactness is greater than $\approx$ 0.2, however the maximum compactness is larger when it is defined in terms of the baryonic mass ($\approx$ 0.43), as opposed to when it is defined in terms of the gravitational mass ($\approx 0.35$). This maximum range of validity is what makes it possible for us to study the evolution of HMNSs on the $f$-$C$ plane (as we show in Sec.~\ref{fc_hmns}), since the collapse only happens when the maximum compactness is reached.

        \subsection{\textbf{\textit{f}}-\textbf{\textit{C}} for hypermassive neutron stars}

            \label{fc_hmns}

            We verify that, considering bulk quantities, the dimensionless $f$-mode frequency Re(${{\tilde{\omega}}^{\rm blk}}_{f}$) correlates well with the compactness ${\tilde{C}}^{\rm blk}_{\rm V}$, resulting in a new universal relation that we refer to as ``bulk $f$-$C$ relation''. We can therefore investigate the evolution of the HMNSs in~\cite{kastaun} on the ${\rm Re}({\tilde{\omega}}^{\rm blk}_{f})-{\tilde{C}}^{\rm blk}_{\rm V}$ plane, as we show in Fig.~\ref{hmns_plot}, for two mass ratios, $q = 0.9$ and $q = 1.0$. This relation allows us to directly compare the gravitational waveform peak frequency of HMNSs ({\it i.e.} the frequency at the peak of the Fourier transform of the postmerger GW signal) with the $f$-mode frequency of NSs with the same compactness, taking the EOS-variation of $\lesssim 3\%$ as a caveat. We see that the deviation of the peak frequency from the universal relation is $\approx$ 73$\%$ for $q = 0.9$ and $\approx$ 78$\%$ for $q = 1.0$, which clearly shows that the universality does not hold for HMNSs.
    
            This result is expected because we did not include rotational or thermal effects in the bulk $f$-$C$ relation. Chakravarti and Andersson~\cite{chakravarti2020} attempted to include these effects in the dimensionless $f$-mode frequency ${\bar{\omega}}_{f}$ of cold and nonrotating NSs by using simple approximations and obtained that ${\bar{\omega}}_{f}$ changes by a factor of $\sim 3$. This factor accounts for the deviation seen in the relation between the premerger tidal deformability and the dimensionless postmerger frequency when compared to the $f$-Love relation for cold and nonrotating NSs. We obtained smaller deviations ($\sim$ 2), considering the bulk $f$-$C$ relation in Fig.~\ref{hmns_plot}. Nonetheless, we note that our independent variable, the bulk compactness of the HMNS, is a postmerger variable in the bulk $f$-$C$ relation, which is not the case in the relation investigated in~\cite{chakravarti2020}.
    
            The inclusion of rotational and thermal effects in cold and nonrotating NSs needs to be done carefully and following what has been found in the current simulations, if we want to recover, {\it e.g.} the peak frequency of the postmerger remnant. For instance, HMNSs can have a core that is slowly rotating, while the maximum angular velocity is reached in the outer parts of the remnant~\cite{kastaun2015, kastaun2, ciolfi2019, kastaun, ciolfi2017, kastaun2017, endrizzi2018, endrizzi2016, hanauske2017}, thus we need an accurate description of the non-uniform rotational profile (see, {\it e.g.}~\cite{iosif2021}). The non-uniform thermal profile (see previously cited works) should also be precisely modeled. 

    \section{Final remarks}

        \label{disc}

        We constructed TOV solutions for PNSs, by using 1D angle-averaged pressure-density relations from the 3D CCSN simulations in~\cite{radice}, considering eight supernova progenitors and the SFHo EOS. We then used relativistic stellar perturbation theory to compute their $f$-mode; their moment of inertia ($I$) and quadrupole moment ($Q$), using the slow rotation approximation; and their tidal deformability (or Love number), using the small tidal deformation approximation. We verified that the $I$-Love-$Q$ and $f$-Love relations for cold NSs are recovered within a postbounce time of $\approx$ 1 s. We showed that these well-established universal relations for cold NSs are approximately valid for PNSs for a postbounce time $\gtrsim 0.5$~s, with a relative difference $\approx 1\% - 10\%$. We were not able to investigate the EOS-variation of these relations.
    
        By using the prescription described in~\cite{kastaun2}, we obtained a new universal relation between the normalized $f$-mode frequency and the compactness of the bulk of cold and nonrotating NSs, with an EOS-variation of $\approx 3\%$. The same bulk quantities for HMNSs were obtained in the BNS merger simulations in~\cite{kastaun}, considering two mass ratios and the SFHo EOS. We confronted the relation between the HMNS quantities from these simulations, namely, the GW peak frequency and the compactness, with the preceding $f$-$C$ relation. As expected, the universal relation was not satisfied by the HMNSs, with a relative difference $\approx 70\% - 80\%$, as rotational and thermal effects are necessary for the description of these objects. The precise modeling of these effects and their impact on, {\it e.g.} the oscillation modes of postmerger remnants is yet to be explored.

        The future detection of a GW signal from a CCSN explosion or a BNS merger could bring valuable information about the macroscopic properties of newly-born PNSs or postmerger remnants. Universal relations would potentially be a suitable tool in the analysis of the upcoming GW data by eliminating the uncertainty caused by the unknown EOS in the parameter estimation. Such relations should be used with care since additional degrees of freedom in the description of, {\it e.g.} PNSs or HMNSs might break their universality. For instance, we showed that some of the well-established universal relations for cold NSs, the $I$-Love-$Q$, $f$-Love, and $f$-$C$ relations, are not valid when rotational and thermal effects (crucial in the modeling of PNSs or HMNSs) are considered.

        Nevertheless, our results suggest that some of these universal relations could still be used with an error $\lesssim 10\%$ ($\lesssim 1\%$) in the early (late) life of a PNS. For example, in the event of a GW detection from a CCSN explosion, the $f$-Love relation could be used to estimate the PNS tidal deformability, which in turn could be used to infer their moment of inertia and quadrupole moment through the $I$-Love-$Q$ relations; further, these measurements could be translated to constraints on the hot EOS. 
        
        Unfortunately, we cannot say the same for the HMNSs, since their $f$-$C$ relation shows high deviations from the NS relation during their short life. Nonetheless, it is possible that HMNSs have universal relations between their quasiperiodic oscillation frequencies and bulk properties. For example, in~\cite{manoharan2021}, the authors proposed relations between the premerger binary tidal deformability and the SMNS compactness. It is known that the GW peak frequency of postmerger remnants is well correlated with the premerger tidal deformability~\cite{breschi2019, gonzalez2022}, so this suggests a relation between the SMNS peak frequency and compactness. The study of such relations for HMNSs, considering its bulk properties, would be possible with the availability of more simulations in the future.

            \renewcommand{\thefigure}{A.1}
            \begin{figure*}
                \centering
                \includegraphics[width=\textwidth]{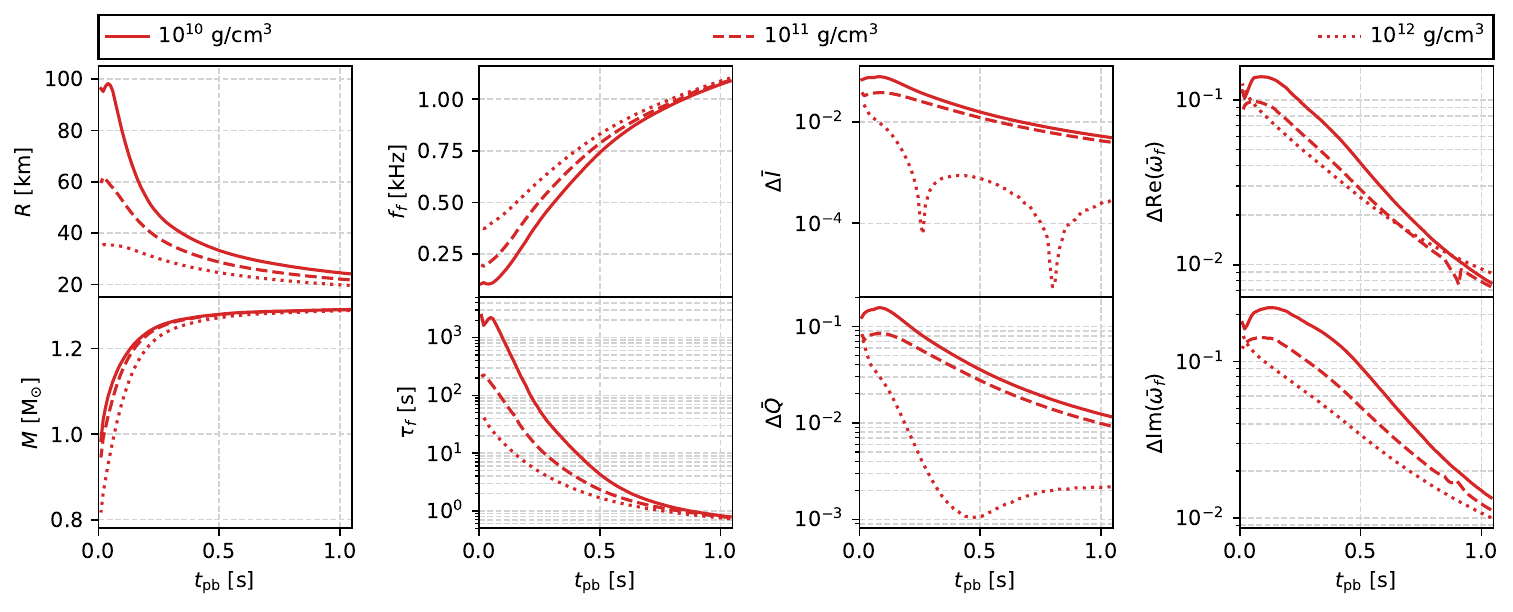}
                \caption{{\it Left panels}: Time evolution of circumferential radius $R$, gravitational mass $M$, $f$-mode oscillation frequency $f_{f}$, and $f$-mode damping time $\tau_{f}$ for three different density cutoffs: $\rho^{\rm srf}_{\rm B} = 10^{10}$ g/cm$^{3}$, $\rho^{\rm srf}_{\rm B} = 10^{11}$ g/cm$^{3}$, and $\rho^{\rm srf}_{\rm B} = 10^{12}$ g/cm$^{3}$, considering the PNS generated by the 9 M$_{\odot}$ progenitor. The differences are more significant at early $t_{\rm pb}$ and decrease as $t_{\rm pb}$ increases. {\it Right panels}: Time evolution of relative difference $\Delta y=|1-y_{\rm PNS}/y_{\rm NS}|$ between PNS observables $y_{\rm PNS}$ and NS observables $y_{\rm NS}$ for three different density cutoffs, where $y \in \{\bar{I},\textrm{ }\bar{Q},\textrm{ }{\rm Re}({\bar{\omega}}_{f}),\textrm{ }{\rm Im}({\bar{\omega}}_{f})\}$. In general, $\Delta y$ decreases as $\rho_{\rm B}^{\rm srf}$ increases. We obtained similar results for the other progenitors.}
                \label{comp_dens}
            \end{figure*}

    \begin{acknowledgments}
    
        The authors thank David Radice and Adam Burrows for providing the data for the 3D CCSN simulations, and helpful discussions; and Wolfgang Kastaun for providing the data for the BNS simulations, and useful e-mail exchanges. V.G. thanks Cole Miller for the informative discussion regarding the radial stability of PNSs. C.C. acknowledges support by NASA under award number 80GSFC21M0002. K.Y. acknowledges support from NSF Grant PHYS-2339969, a Sloan Foundation Research Fellowship and the Owens Family Foundation. 

    \end{acknowledgments}
  
    \begin{appendices}

        \section{Different density cutoffs for protoneutron stars}

            \counterwithin{figure}{section}
            \counterwithin{table}{section}

            \label{apA}

            The PNS radius is not uniquely defined, since the boundary between the ``interior'' and the ``exterior'' of the star is not clear. The ``surface'' is usually defined by a baryonic mass density cutoff $\rho^{\rm srf}_{\rm B}$ that needs to be carefully chosen. For example, Torres-Forné {\it et al.}~\cite{torres2018} points out that considering the neutrinosphere ($\rho^{\rm srf}_{\rm B} \sim 10^{12}$ g/cm$^{3}$) as a proxy for the radius can underestimate the PNS size, and thus considers $\rho^{\rm srf}_{\rm B} = 10^{11}$ g/cm$^{3}$ as a more realistic cutoff. This is indeed the regular choice in the literature (see~\cite{muller2013, takiwaki2014, sotani2016, torres2019}). In this work, we follow~\cite{morozova}, and our main results are presented with  $\rho^{\rm srf}_{\rm B} = 10^{10}$ g/cm$^{3}$ (see, {\it e.g.} Fig.~\ref{fig02}), which was also adopted in~\cite{sotani2016, sotani2017}.
                
            The definition of the ``surface'' has an impact on the macroscopic properties of the PNSs. For instance, the $f$-mode frequency of the PNSs depends on the outer boundary condition for the eigenvalue problem. In particular, Sotani {\it et al.}~\cite{sotani2019} showed that the oscillation frequencies are affected by the outer boundary, characterized either by constant density surfaces or by the shock radius (as done previously in~\cite{torres2018}). Here, we compare how different density cutoffs affect our results. We obtain TOV solutions for three density cutoffs: $\rho^{\rm srf}_{\rm B} = 10^{10}$ g/cm$^{3}$, $\rho^{\rm srf}_{\rm B} = 10^{11}$ g/cm$^{3}$, and $\rho^{\rm srf}_{\rm B} = 10^{12}$ g/cm$^{3}$. These choices are in accordance with most of the previously mentioned works and comprise a broad range for the PNS radius.

            We show results for the circumferential radius, the gravitational mass, and the $f$-mode, for the PNS generated by the 9 M$_{\odot}$ progenitor, in the left panels of Fig.~\ref{comp_dens}. As $\rho^{\rm srf}_{\rm B}$ increases, the PNS radius and mass decrease for all $t_{\rm pb}$, as expected. This difference is smaller at late $t_{\rm pb}$ and it becomes even less important as $t_{\rm pb}$ increases further since the PNS is slowly cooling. As for the $f$-mode, we note that the frequency increases and the damping time decreases for all $t_{\rm pb}$, as a consequence of the change in the radius and the mass, or, to be more precise, the increase in the average density. Similarly, as $t_{\rm pb}$ increases, the different choices for the density cutoff do not affect the $f$-mode significantly. 

            \renewcommand{\thefigure}{B.1}
            \begin{figure*}
                \centering
                \includegraphics[width=\textwidth]{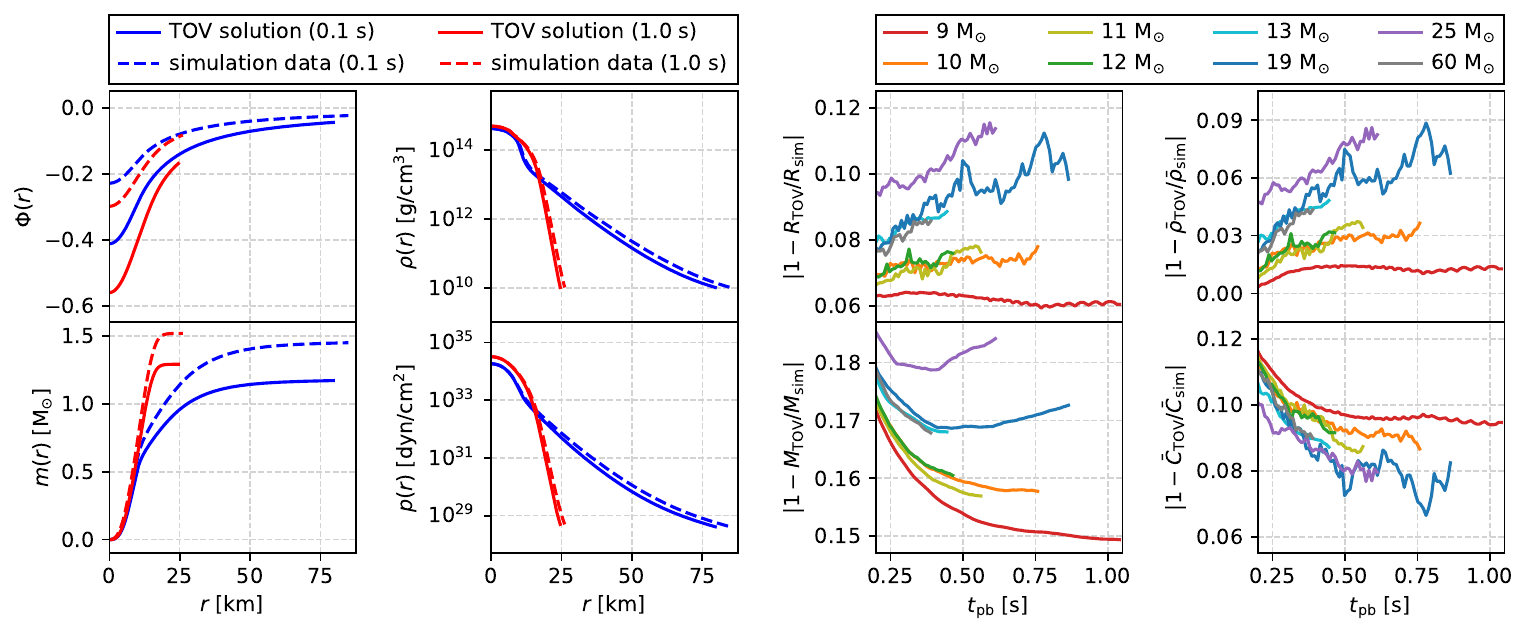}
                \caption{{\it Left panels}: Radial profiles for the gravitational potential $\Phi(r)$, gravitational mass $m(r)$, total mass density $\rho(r)$, and pressure $p(r)$ for our TOV solutions and the simulation data. The profiles are for the PNSs generated by the 9 M$_{\odot}$ progenitor at $t_{\rm pb} = 0.1$ s and $t_{\rm pb} = 1.0$ s. We note that the $\Phi(r)$ profiles for our TOV solutions are deeper, slightly modifying the $\rho(r)$ and $p(r)$ profiles, and thus affecting the $m(r)$ profile. {\it Right panels}: Relative differences between PNS observables from TOV solutions $y_{\rm TOV}$ and simulation data $y_{\rm sim}$, where $y \in \{R,\textrm{ }M,\textrm{ }\bar{\rho}=\sqrt{M/R^{3}},\textrm{ }\bar{C}=M/R\}$, and for $t_{\rm pb} \gtrsim 0.2$ s. In general, $y_{\rm TOV} < y_{\rm sim}$, considering the eight different progenitors (see Table~\ref{tab_pns}).} 
                \label{comp}
            \end{figure*}

            The $f$-Love relation at late $t_{\rm pb}$ is not severely affected by $\rho^{\rm srf}_{\rm B}$, as we show in the right panels of Fig.~\ref{comp_dens}, where we also include the relative error for the $I$-Love and $Q$-Love relations. We note that the relative difference between the PNS relations and the NS relations decreases as $\rho_{\rm B}^{\rm srf}$ increases, {\it i.e.} universality is recovered earlier. In particular, for $\rho_{\rm B}^{\rm srf} = 10^{12}$ g/cm$^{3}$, the PNS $I$-Love relation crosses the NS $I$-Love relation (a similar behavior is seen for the imaginary part of the $w$-Love relation for $\rho_{\rm B}^{\rm srf} = 10^{10}$ g/cm$^{3}$, see Appendix \ref{apD}). We also performed this analysis for the other progenitors and the results are similar. 

            \renewcommand{\thefigure}{B.2}
            \begin{figure}[b]
                \centering
                \includegraphics[width=0.49\textwidth]{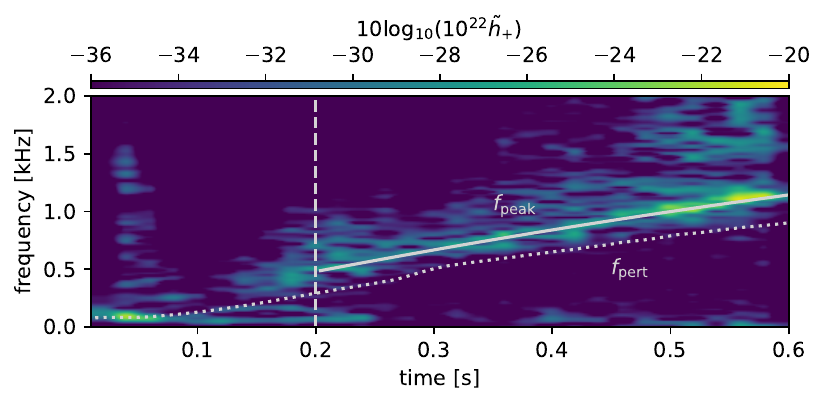}
                \caption{GW spectrogram for the 25 M$_{\odot}$ progenitor. The $f$-mode frequency ($f_{\rm pert}$, computed using perturbation theory) is lower than the simulation frequency ($f_{\rm peak}$, obtained from the short-time Fourier transform), and we attribute this difference to the approximations used in the simulations in~\cite{radice} (see main text).}
                \label{spec_sev}
            \end{figure}

        \section{TOV solutions \textbf{\textit{vs.}} simulation data for protoneutron stars}

            \counterwithin{figure}{section}
            \counterwithin{table}{section}

            \label{apB}

            The treatment of gravity in the CCSN simulations in~\cite{radice} is not fully general-relativistic. The 3+1 formalism was used and the lapse function $\alpha = {\rm exp}(\Phi_{\rm eff})$ was computed in terms of an effective relativistic potential $\Phi_{\rm eff}$ (see Case A in~\cite{marek2006}) that serves as an approximation for the TOV potential. Moreover, the conformal flatness condition was adopted and the conformal factor $\psi$ was fixed to $1$. It is then expected that the radial profiles for our TOV solutions should differ from the radial profiles from the simulations since we are taking the time-dependent $p$~{\it vs.}~$\rho$ relation as an effective EOS and solving the general-relativistic equations. As $\psi = 1$, the isotropic radial coordinate $\bar{r}$ is equivalent to the Schwarzschild radial coordinate $r$, and thus it is straightforward to perform this comparison. We present these results in the left panels of Fig.~\ref{comp}, where we show the gravitational potential $\Phi(r) = \ln\alpha(r)$, the gravitational mass $m(r)$, the total mass density $\rho(r)$, and the pressure $p(r)$ profiles at $t_{\rm pb} = 0.1$ s and $t_{\rm pb} = 1.0$ s for the PNS generated by the 9 M$_{\odot}$ progenitor.

            \renewcommand{\thefigure}{C.1}
            \begin{figure*}
                \centering
                \includegraphics[width=\textwidth]{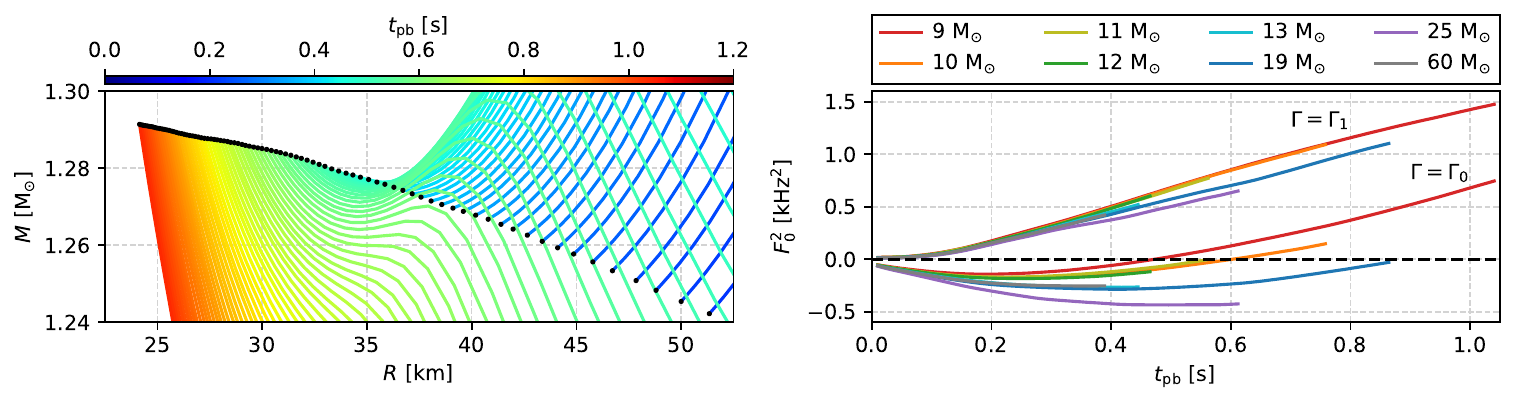}
                \caption{{\it Left panel}: $M(t_{\rm pb})$ {\it vs.} $R(t_{\rm pb})$ relation for the PNS generated by the 9 M$_{\odot}$ progenitor. The black points show the stars that we are actually considering in this work, since their central density is the same as the one in the simulation data. We extend these configurations to lower central densities, and the colorful curves show the slope of the mass-radius relation for each time slice. We note that, for $t_{\rm pb} \gtrsim 0.5$ s, the black points are in stable branches ({\it i.e.} the mass is increasing and the radius is decreasing). {\it Right panel}: Time evolution of the fundamental mode squared $F^{2}_{0}$ for the radial oscillations of PNSs generated by eight different progenitors (see Table~\ref{tab_pns}). We note that, for the 9 M$_{\odot}$ progenitor: (i) if $\Gamma$ = $\Gamma_{0}$, $F_{0}^{2} > 0$ for $t_{\rm pb} \gtrsim 0.5$ s, which agrees with the qualitative analysis using the turning point principle; (ii) if $\Gamma = \Gamma_{1}$, $F_{0}^{2} > 0$ for all $t_{\rm pb}$, thus breaking the validity of the maximum-mass criterion (see main text). We obtained similar results for the other progenitors.}
                \label{fig03}
            \end{figure*}

            \renewcommand{\thefigure}{C.2}
            \begin{figure}[b]
                \centering
                \includegraphics[width=0.49\textwidth]{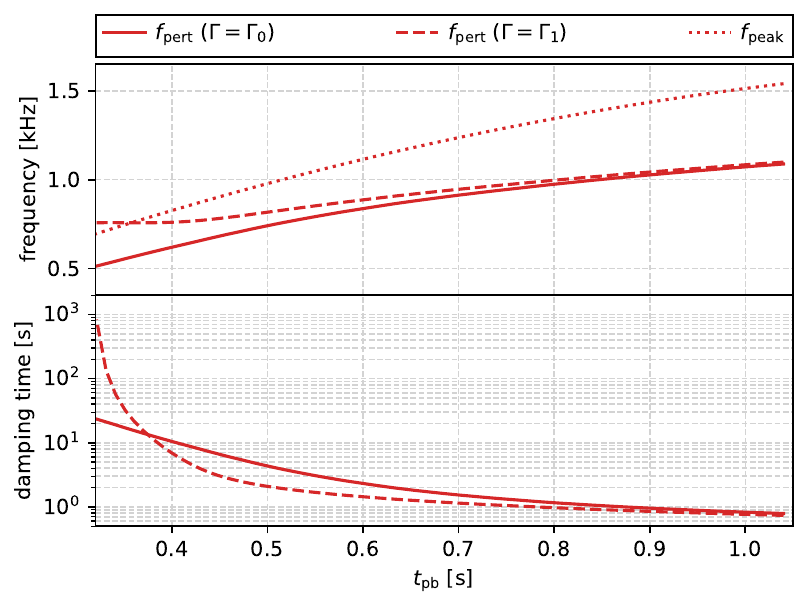}
                \caption{Time evolution of the $f$-mode frequency (upper panel) and damping time (lower panel) for the 9 M$_{\odot}$ progenitor, computed with perturbation theory, considering $\Gamma = \Gamma_{0}$ and $\Gamma = \Gamma_{1}$ (see main text). For comparison, we also show the GW frequency obtained from the spectrogram, with a short-time Fourier transform (see Figs.~\ref{comp_rod_afle}~and~\ref{spec_sev}). We obtained similar results for the other progenitors.}
                \label{fmode_g1}
            \end{figure}

            We note that, as expected, the potential $\Phi(r)$ is deeper in our TOV solutions. This difference in the potential is responsible for the slight modification in the profiles of the pressure $p(r)$ and the density $\rho(r)$ (relative to the simulation data, although $\bar{r} = r$), which, in turn, affects the profile of the mass $m(r)$. In particular, we note that the total gravitational mass, $M=m(r=R)$, where $R$ is the circumferential radius (defined by a density cutoff of $\rho_{\rm B}^{\rm srf} = 10^{10}$ g/cm$^{3}$, see Sec.~\ref{tov_sol}), is lower in our TOV solutions. In the right panels of Fig.~\ref{comp}, we compare results for different observables for the PNSs from the eight progenitors, considering $t_{\rm pb} \gtrsim 0.2$ s (this is the minimum time such that we perform the comparison for the $f$-mode frequency between the TOV solutions and the simulation data, following Afle {\it et al.}~\cite{chaitanya2023}, see Fig.~\ref{comp_rod_afle}). We show the relative difference $|1-y_{\rm TOV}/y_{\rm sim}|$ between PNS observables computed from our TOV solutions $y_{\rm TOV}$ and from the simulation data $y_{\rm sim}$, where $y \in \{R,\textrm{ }M,\textrm{ }\bar{\rho},\textrm{ }\bar{C}\}$; here, $\bar{\rho} = \sqrt{M/R^{3}}$ is the average density and $\bar{C} = M/R$ is the compactness. We note that the results from our TOV solutions (i) for the radius and the compactness are lower by $\approx 6-12\%$; (ii) for the total mass are lower by $\approx15-19\%$; (iii) for the average density are lower by $\approx 1-9\%$. Regardless of these differences, we use these TOV solutions as backgrounds for perturbations, and compute, {\it e.g.} their $f$-mode frequency. 

            The GW signal from the CCSN simulations in~\cite{radice} was obtained through the quadrupole formula (see, {\it e.g.}~\cite{finn1990}), that gives an approximation to the amplitude of the signal. Thus, we should expect that the time-dependent frequencies of these GW signals should differ from the frequencies computed using perturbation theory. We present this comparison in Fig.~\ref{spec_sev}, where we show the spectrogram for the GW signal of the PNS generated by the 25 M$_{\odot}$ progenitor, which has the most energetic signal~\cite{radice}. We extracted the frequency of the GW signal, which we denote as $f_{\rm peak}$, through a short-time Fourier transform analysis, following~\cite{chaitanya2023}. The peak frequency $f_{\rm peak}$ is characterized by the high-power ``track'' on the time-frequency plane of Fig.~\ref{spec_sev}. We denote the frequency obtained through perturbation theory as $f_{\rm pert}$. We see that $f_{\rm pert}$ as a function of the postbounce time is lower than $f_{\rm peak}$, with a relative difference $\lesssim 40\%$ (see lower left panel of Fig.~\ref{comp_rod_afle} for more details). We attribute these differences to the approximations that were adopted in the simulations in~\cite{radice} for the description of the spacetime and the estimation of the GW signal, which produced simulated PNSs that are less compact than our TOV solutions (see Fig.~\ref{comp}).

        \section{Radial stability of protoneutron stars and the adiabatic index}

            \counterwithin{figure}{section}
            \counterwithin{table}{section}

            \label{apC}

            \renewcommand{\thefigure}{D.1}
            \begin{figure*}
                \centering
                \includegraphics[width=\textwidth]{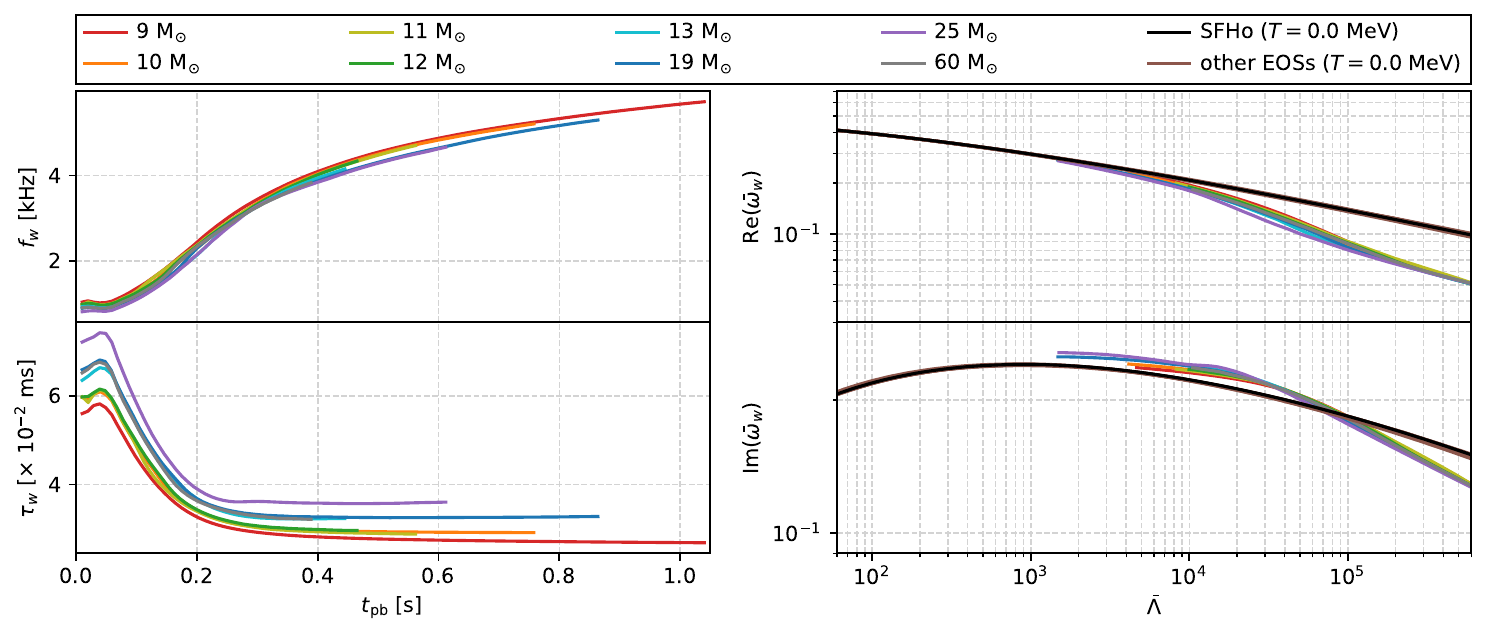}
                \caption{{\it Left panel}: Time evolution of the oscillation frequency $f_{w}$ and the damping time $\tau_{w}$ of the $w$-mode of PNSs generated by eight different progenitors (see Table~\ref{tab_pns}). Similarly to the $f$-mode case (see Fig.~\ref{fig02}), the frequencies are lower while the damping times are (slightly) longer than those for cold NSs ($f^{\rm NS}_{w} \approx 5-10$ kHz and $\tau^{\rm NS}_{w} \approx 0.01-0.05$ ms). {\it Right panel}: Time evolution of the $w$-Love relation. We consider $\bar{\Lambda}$ as the independent variable and show its correlation with Re($\bar{\omega}_{w}$) and Im($\bar{\omega}_{w}$). For reference, we include the relations for cold NSs, considering ten zero-temperature EOSs (as in Fig.~\ref{ilq_err}). Similarly to the case for $I$-Love-$Q$ or $f$-Love, the PNS relations approach the cold NS ones as the postbounce time $t_{\rm pb}$ increases (from high $\bar{\Lambda}$ to low $\bar{\Lambda}$).}
                \label{wmode}
            \end{figure*}

            For cold NSs, the mass-radius relation has a maximum point, which corresponds to the maximum mass $M_{\rm max}$. NSs become radially unstable when $M(\rho_{0})>M_{\rm max}$, such that $\partial{M}/\partial{\rho_{0}} < 0$, where $\rho_{0}$ is the central total mass density. This is generally known as the ``maximum-mass'' criterion or the ``turning-point'' principle (see, {\it e.g.}~\cite{hadzic2021}). The mass-radius relation can have multiple maximum points, {\it i.e.} more than one stable branch, when the EOS have first-order phase transitions~\cite{glendenning2000}; further, it can have extended branches of stability when the phase conversion is slow~\cite{lugones}. The existence of multiple stable branches or extended branches in the mass-radius relation has already been shown to severely affect well-established universal relations for cold NSs ({\it e.g.} $f$-Love~\cite{ranea2023} and $w$-Love~\cite{ranea2022}).

            For PNSs, $\rho_{0}$ changes as $t_{\rm pb}$ increases; thus, we cannot interpret the mass-radius relation as for cold NSs. Naïvely, we can treat $t_{\rm pb}$ as a parameter and obtain mass-radius relations for our PNSs, namely we can construct multiple TOV solutions  for a fixed $t_{\rm pb}$ with various $\rho_{0}$ that are smaller than that used in the simulation and obtain a ``mass-radius relation'' for each $t_{\rm pb}$. We illustrate this procedure in the left panel of Fig.~\ref{fig03}, for the PNS generated by the 9 M$_{\odot}$ progenitor, where the black points represent the stars with $\rho_{0}$ given by the data and the colorful curves represent the ``mass-radius relations'' for these stars. We note that the configurations are radially stable only for $t_{\rm pb} \gtrsim 0.5$ s.

            We can verify the maximum-mass criterion through an analysis of the radial oscillations of the PNSs by checking the sign of their fundamental radial mode frequency squared $F^{2}_{0}$. If $F_{0}^{2} > 0$, the configurations are radially stable, since $F^{2}_{0} < F^{2}_{1} < \hdots$, by definition~\cite{chandrasekhar1964}. Gondek {\it et al.}~\cite{gondek} already pointed out that the maximum-mass criterion is approximately valid for PNSs, regardless of thermal and neutrino trapping effects. These effects are encoded in the adiabatic index $\Gamma$, a quantity that relates the perturbed pressure and density in the oscillating star (see, {\it e.g.}~\cite{haensel2002}). In general, we have $\Gamma \equiv (\partial{\ln{p}}/\partial{\ln{\rho_{\rm B}}})_{\rm C}$,  where ``C'' stands for some condition to be applied on the derivative. In equilibrium, this definition reduces to $\Gamma_{0} \equiv (\textrm{d}\ln{p}/\textrm{d}\ln{\rho_{\rm B}})_{\rm EOS}$, where the derivative is taken from the $p$~{\it vs.}~$\rho_{\rm B}$ relation ({\it i.e.}, the EOS) of the unperturbed background; or to $\Gamma_{1}\equiv(\partial{\ln{p}}/\partial{\ln{\rho_{\rm B}}})_{s,\{Y_i\}}$, where the derivative is taken at constant entropy per baryon $s$ and particle abundances $\{Y_i\}$.
            
            We consider these two cases, where we obtain $\Gamma_{0}$ from the background effective EOS (see Footnote~\ref{note}), and $\Gamma_{1}$ from the simulations. In the right panel of Fig.~\ref{fig03}, we confirm that, for the PNS generated by the 9 M$_{\odot}$ progenitor, with $\Gamma=\Gamma_0$, the configurations are indeed radially stable only for $t_{\rm pb} \gtrsim 0.5$ s. We obtained similar results for the other progenitors. This radial instability for early times is likely to be not physical since the PNS is losing energy through neutrino emission and this process could act as a dissipation mechanism, especially it would prevent the exponential growth of the radial oscillations. 

            In fact, by taking the effects of the perturbed EOS into account, {\it i.e.} when considering $\Gamma = \Gamma_{1}$, we obtain that the PNSs are radially stable for all times, as we show in the right panel of Fig.~\ref{fig03}. Thus, the temperature and composition gradients encoded in $\Gamma_{1}$ are crucial for the determination of the radial stability of the PNSs, and the maximum-mass criterion is not applicable. The results in the right panel of Fig.~\ref{fig03} are for the PNS radius defined by the density cutoff $\rho^{\rm srf}_{\rm B} = 10^{10}$ g/cm$^{3}$, but we also computed the frequencies for different density cutoffs ($\rho^{\rm srf}_{\rm B} = 10^{11}$ g/cm$^{3}$ and $\rho^{\rm srf}_{\rm B} = 10^{12}$ g/cm$^{3}$, see Appendix~\ref{apB}) and the results are similar.

            \renewcommand{\thefigure}{E.1}
            \begin{figure*}
                \centering
                \includegraphics[width=\textwidth]{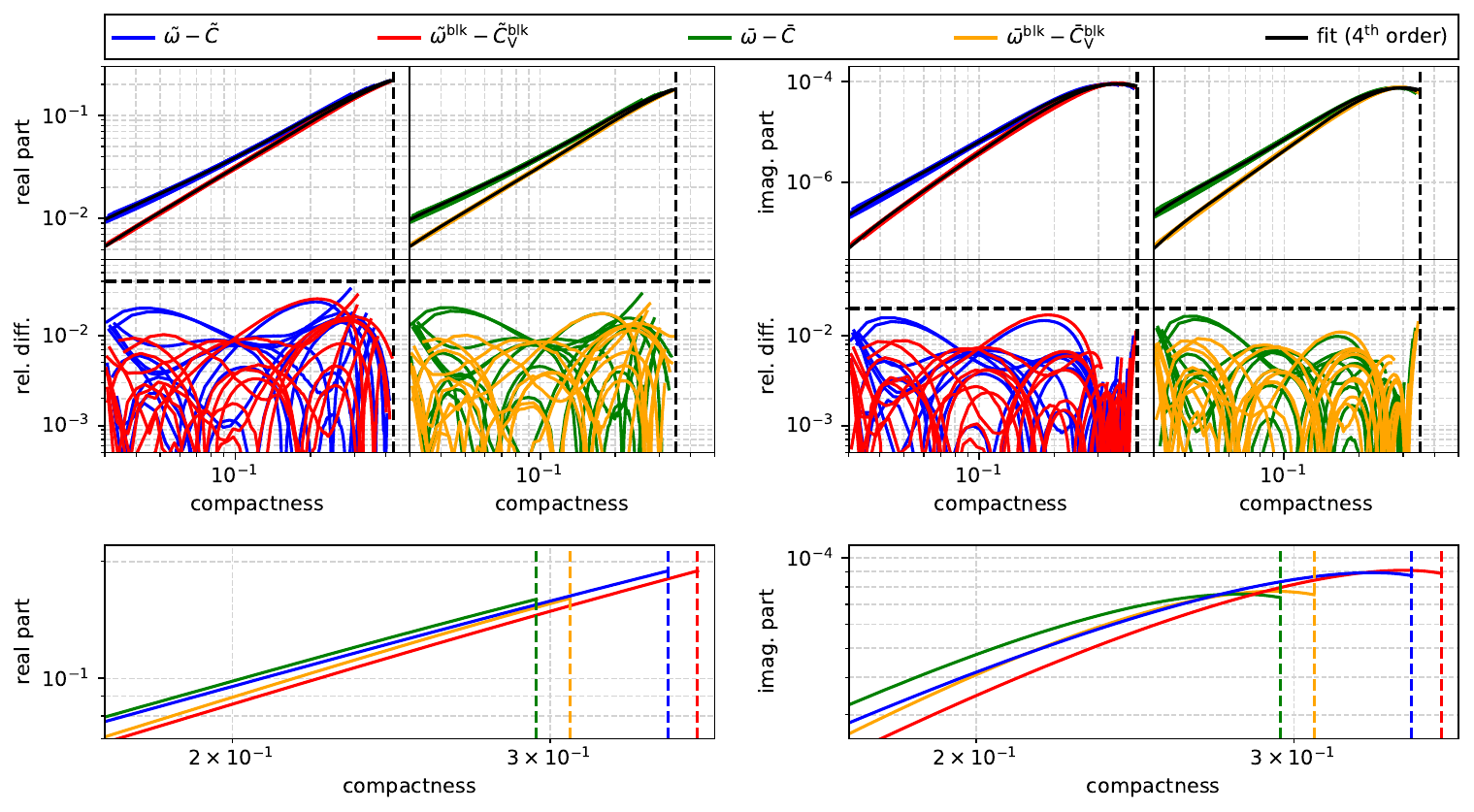}
                \caption{{\it Upper panels}: NS and bulk $f$-$C$ relations for different definitions (see main text). The bulk relations are below the NS ones (except for the imaginary part, when the compactness is close to its maximum). The maximum compactness (see vertical dashed lines) for the tilde-defined (defined in terms of the baryonic mass) quantities is $\approx 0.43$, and for the bar-defined (defined in terms of the gravitational mass) quantities is $\approx 0.35$. The relative difference $|1-y_{\rm fit}/y|$ between $y_{\rm fit}$ and $y$ is $\lesssim 4\%$ for the real part relations and $\lesssim 2\%$ for the imaginary part relations (see the horizontal dashed lines), where $y$ is a dependent variable. {\it Lower panels}: $f$-$C$ relations corresponding to the different definitions (see main text), for the SFHo EOS. For this compactness range (that is similar to the one in Fig.~\ref{hmns_plot}), the relations are very similar, although the maximum range of their validity is significantly different (see vertical dashed lines).}
                \label{fc_plot_sev}
            \end{figure*}

            We performed a similar analysis for the nonradial $f$-mode frequency, namely we compared the frequencies and damping times for $\Gamma=\Gamma_{0}$ and $\Gamma=\Gamma_{1}$. As opposed to the fundamental radial mode, we obtained that the results agree for later postbounce times, as we show in Fig.~\ref{fmode_g1} for the 9 M$_{\odot}$ progenitor (which corresponds to the longest simulation available, see Table~\ref{tab_pns}). Thus, the approximation $\Gamma_{1} \approx \Gamma_{0}$ does not affect our results regarding the time evolution of the universal relations (see Sec.~\ref{ilq}). For earlier times ($t_{\rm pb} \lesssim 0.32$ s), we were not able to compute the $f$-mode in terms of the Cowling classification, since its radial eigenfunction starts having nodes (similar to what has been found in~\cite{morozova, rodriguez2023}).

        \section{\textbf{\textit{w}}-Love for protoneutron stars}

            \counterwithin{figure}{section}
            \counterwithin{table}{section}

            \label{apD}

            The GW spectrum of the CCSN simulations in~\cite{radice} at late $t_{\rm pb}$ shows a clear signature (see Fig.~\ref{spec_sev}). At early $t_{\rm pb}$ ($\lesssim 0.4$ s), we can describe the spectrum using the gravity ($g$-) and pressure ($p$-) modes of the PNS (considering the Cowling classification; however, this is not the only possible classification for the oscillation modes, see~\cite{rodriguez2023}), but at late $t_{\rm pb}$ ($\gtrsim 0.4$ s), we can characterize the narrow ``track'' in the GW spectrograms using the fundamental ($f$-) mode. These ($f$-, $p$-, and $g$-) modes are associated with perturbations of the remnant PNS fluid, which are coupled to the perturbations of the spacetime. 
            
            As a result of the highly asymmetric collapse, spacetime deformations are expected, and purely spacetime ($w$-) modes can be excited~\cite{andersson1996}. These modes are mostly associated with oscillations of the spacetime and are barely coupled to any fluid motion. $w$-modes can be classified as curvature, interface, or trapped modes; the curvature modes are the most relevant $w$-modes in astrophysical applications~\cite{kokkotas} and for cold NSs, their frequency is $\approx$ 5 $-$ 10 kHz and their damping time is $\approx$ 0.01 $-$ 0.05 ms, which are, respectively, higher and shorter, compared to any of the $f$-, $p$-, or $g$-modes. Unfortunately, $w$-modes are yet to be identified in the GW spectrum of CCSN simulations. Nevertheless, it is instructive to see how their frequencies compare to the frequencies of the fluid modes of PNSs and how the well-established universal relations for cold NSs are affected by thermal effects on the effective EOS.

            \renewcommand{\thetable}{E.1}
            \begin{table*}
                \centering
                \begin{ruledtabular}
                    \begin{tabular}{ccccccccc}
                        $10^{y}$ & Re(${\tilde{\omega}}_{f}^{\rm blk}$) & Im(${\tilde{\omega}}_{f}^{\rm blk}$) & Re(${\bar{\omega}}_{f}^{\rm blk}$) & Im(${\bar{\omega}}_{f}^{\rm blk}$) & Re(${\tilde{\omega}}_{f}$) & Im(${\tilde{\omega}}_{f}$) & Re(${\bar{\omega}}_{f}$) & Im(${\bar{\omega}}_{f}$) \\
                        \hline
                        $10^{x}$ & ${\tilde{C}}_{\rm V}^{\rm blk}$ & ${\tilde{C}}_{\rm V}^{\rm blk}$ & ${\bar{C}}_{\rm V}^{\rm blk}$ & ${\bar{C}}_{\rm V}^{\rm blk}$ & $\tilde{C}$ & $\tilde{C}$ & $\bar{C}$ & $\bar{C}$ \\
                        \hline \hline
                        $c_{0}$ & $-6.253\times10^{-1}$ & $-6.411$ & $-9.036\times10^{-1}$ & $-9.041$ & $-6.870\times10^{-1}$ & $-6.520$ & $-1.046$ & $-9.286$ \\
                        \hline
                        $c_{1}$ & $-9.937\times10^{-1}$ & $-1.295\times10^{1}$ & $-2.255$ & $-2.375\times10^{1}$ & $-1.272$ & $-1.313\times10^{1}$ & $-2.830$ & $-2.428\times10^{1}$ \\
                        \hline
                        $c_{2}$ & $-3.780$ & $-2.275\times10^{1}$ & $-5.670$ & $-3.834\times10^{1}$ & $-4.133$ & $-2.266\times10^{1}$ & $-6.450$ & $-3.856\times10^{1}$ \\
                        \hline
                        $c_{3}$ & $-2.515$ & $-1.407\times10^{1}$ & $-3.697$ & $-2.367\times10^{1}$ & $-2.803$ & $-1.407\times10^{1}$ & $-4.257$ & $-2.380\times10^{1}$ \\
                        \hline
                        $c_{4}$ & $-6.089\times10^{-1}$ & $-3.257$ & $-8.757\times10^{-1}$ & $-5.421$ & $-6.649\times10^{-1}$ & $-3.208$ & $-9.951\times10^{-1}$ & $-5.390$ \\
                        \hline \hline
                        $\Delta y^{\rm rms}$ & $1.001\times10^{-2}$ & $4.189\times10^{-3}$ & $8.132\times10^{-3}$ & $3.845\times10^{-3}$ & $1.092\times10^{-2}$ & $4.369\times10^{-3}$ & $9.556\times10^{-3}$ & $3.993\times10^{-3}$ \\
                        \hline
                        $\Delta y^{\rm max}$ & $2.825\times10^{-2}$ & $1.693\times10^{-2}$ & $2.258\times10^{-2}$ & $1.385\times10^{-2}$ & $3.297\times10^{-2}$ & $1.574\times10^{-2}$ & $2.904\times10^{-2}$ & $1.640\times10^{-2}$ \\
                    \end{tabular}
                \end{ruledtabular}
                \caption{Same as Table~\ref{table_coeff_1} but for the different $f$-$C$ relations shown in Fig.~\ref{fc_plot_sev} and $n = 4$.}
                \label{table_coeff_2}
            \end{table*}

            Following Sotani {\it et al.}~\cite{sotani2017}, we focus on the fundamental axial $w$-mode, the first curvature mode, which we simply refer to as the $w$-mode. Sotani {\it et al.}~\cite{sotani2017} proposed that, for PNSs, the relation between the frequency of the $w$-mode (multiplied by the PNS radius) and the compactness is universal. This relation is referred to as the ``$w$-$C$ relation'', and was already studied for cold NSs (see,~{\it e.g.}~\cite{andersson1998, tsui2005}). Blázquez-Salcedo {\it et al.}~\cite{salcedo2013} investigated phenomenological relations for the axial $w$-modes, and Mena-Fernández and González-Romero~\cite{fernandez2019} related the $w$-mode to the tidal deformability, to which we refer to as the ``$w$-Love relation'' (the relation between the $w$-mode and the quadrupole moment was studied in~\cite{benitez2021}). Here, we focus on the $w$-Love relation.

            The eigenvalue problem for axial perturbations is simpler than the one for polar perturbations. We solve the perturbation equation derived by~\cite{chandra}. We impose regularity of the solution at the center, and match it to the Regge-Wheeler function and its derivative at the surface. Then, we use a shooting method to find the complex frequency that gives us an outgoing wave solution for the Regge-Wheeler equation at infinity, using the continued fraction method (see Appendix B of~\cite{sotani}).
 
            We show results for the frequencies and damping times, and the $w$-Love relation in Fig.~\ref{wmode}. We see that, similarly to the $f$-mode case, the frequencies and damping times for the $w$-mode of the PNSs are, respectively, lower and longer than those for cold NSs. The damping times, in particular, are only slightly longer, as opposed to the $f$-mode damping times that are longer than the cold NS ones by up to four orders of magnitude (see right panel of Fig.~\ref{fig02}). In principle, these lower frequencies could facilitate the detection of the $w$-mode in the future. Nonetheless, it is still unclear if the contribution of this mode to the GW signal from a CCSN is relevant enough in terms of radiated energy~\cite{sotani2017}. Moreover, we note that the PNS relations approach the cold NS relations as the postbounce time $t_{\rm pb}$ increases (from high $\bar{\Lambda}$ to low $\bar{\Lambda}$), similarly to the case for $I$-Love-$Q$ or $f$-Love. In particular, we note that the imaginary part of the $w$-Love relation for the PNSs first crosses and then approach the cold NS relation. 

        \section{Different \textbf{\textit{f}}-\textbf{\textit{C}} relations for cold and nonrotating neutron stars}

            \counterwithin{figure}{section}
            \counterwithin{table}{section}

            \label{apE}

            The relation between the $f$-mode complex frequency and the compactness of cold and nonrotating NSs, the $f$-$C$ relation, was proposed by Tsui and Leung~\cite{tsui2005}. In Sec.~\ref{sec:HMNS}, we proposed a new universal relation between the $f$-mode frequency and the compactness, using bulk quantities of cold and nonrotating NSs (the bulk $f$-$C$ relation). On the one hand, the compactness of the bulk is defined in terms of the baryonic mass and the volumetric radius of the bulk (${\tilde{C}}^{\rm blk}_{\rm V} \equiv M^{\rm blk}_{\rm B}/R^{\rm blk}_{\rm V}$, see Sec.~\ref{sec:HMNS})~\cite{kastaun2}. On the other hand, the usual compactness is defined in terms of the gravitational mass and the circumferential radius of the NS ($\bar{C}\equiv M/R$). In the same way, the dimensionless $f$-mode frequency of the bulk is defined in terms of the baryonic mass (${\tilde{\omega}}^{\rm blk} \equiv M^{\rm blk}_{\rm B}\omega^{\rm blk}$, see Sec.~\ref{sec:HMNS}), while the dimensionless $f$-mode frequency of the NS is defined in terms of the gravitational mass ($\bar{\omega} \equiv M\omega$). When comparing the bulk $f$-$C$ relation with the conventional $f$-$C$ relation in the literature, we have to take these differences into account.

            We define the following quantities: ${\bar{C}}^{\rm blk}_{\rm V} \equiv M^{\rm blk}/R^{\rm blk}_{\rm V}$ and ${\bar{\omega}}^{\rm blk} \equiv M^{\rm blk}\omega^{\rm blk}$, and $\tilde{C}\equiv M_{\rm B}/R$ and $\tilde{\omega} \equiv M_{\rm B}\omega$, where $M^{\rm blk}$ is the gravitational mass of the bulk and $M_{\rm B}$ is the baryonic mass of the NS. Thus, we can compare ${\bar{\omega}}^{\rm blk}-{\bar{C}}_{\rm V}^{\rm blk}$ with $\bar{\omega}-\bar{C}$ (the usual $f$-$C$ relation), and ${\tilde{\omega}}^{\rm blk}-{\tilde{C}}_{\rm V}^{\rm blk}$ (the bulk $f$-$C$ relation) with $\tilde{\omega}-\tilde{C}$. We present this comparison in the upper panels of Fig.~\ref{fc_plot_sev}. We are only showing results for radially stable stars. We note that: (i) for the same compactness, the real and imaginary parts of the dimensionless complex frequency of the bulk are smaller than the NS ones (except for the imaginary part, when the compactness is close to the maximum one, see lower panels of Fig.~\ref{fc_plot_sev}); (ii) the different definitions for the dimensionless complex frequency and the compactness, with respect to the baryonic mass (tilde-defined quantities) or the gravitational mass (bar-defined quantities), affect the maximum compactness: for the tilde-defined quantities, it is $\approx 0.43$, and for the bar-defined quantities it is $\approx 0.35$; (iii) the relative difference $|1-y_{\rm fit}/y_{\rm num}|$ between fourth-order fit values $y_{\rm fit}$ and numerical values $y_{\rm num}$ is $\lesssim 4\%$ for the real part relations and $\lesssim 2\%$ for the imaginary part relations. The fitting coefficients are shown in Table~\ref{table_coeff_2}. In general, the different relations are similar when the compactness is greater than $\approx$ 0.2, and the maximum compactness is larger when it is defined in terms of the baryonic mass.

            The lower panels of Fig.~\ref{fc_plot_sev} are zoomed-in versions of the upper panels, showing the relations for the SFHo EOS only. The range for the compactness is similar to the one in Fig.~\ref{hmns_plot}, where we show the evolution of the HMNSs. We note that, although the relations are similar, the maximum compactnesses for the tilde-defined and bar-defined relations are significantly different. In an attempt to include rotational or thermal effects in these relations, we should take these differences into account, and consider not only the change in the frequency but also the changes in mass and radius.

    \end{appendices}
    
    \bibliography{references}

\end{document}